\documentclass[twoside,leqno,twocolumn]{article}

\usepackage{ltexpprt}

\usepackage{algorithm}
\usepackage[noend]{algpseudocode}
\usepackage{amsmath,amsfonts}
\usepackage{booktabs}
\usepackage{csquotes}
\usepackage{graphicx}
\usepackage{multirow}
\usepackage[autolanguage]{numprint}
\usepackage{subcaption}
\usepackage{xcolor}
\usepackage{xspace}
\usepackage{url}
\newcommand{\ie}{i.\,e.,\xspace}
\newcommand{\eg}{e.\,g.,\xspace}
\newcommand{\wrt}{w.\,r.\,t.\xspace}

\newcommand{\etal}{et al.\xspace}
\newcommand{\erdosr}{Erd\H{o}s R\'enyi\xspace}
\newcommand{\ba}{Barab\`asi-Albert\xspace}
\newcommand{\speed}[1]{$\numprint{#1}\times$}

\DeclareMathOperator*{\argmax}{arg\,max}

\newcommand{\bcfn}{\mathrm{BC}}
\newcommand{\gbcfn}{\mathrm{GBC}}
\newcommand{\katzfn}{\mathrm{KC}}
\newcommand{\edfn}{\mathrm{ED}}
\newcommand{\gedfn}{\mathrm{GED}}
\newcommand{\gccfn}{\mathrm{GCC}}

\newcommand{\gedSpeedupGBCkfive}{52.6}
\newcommand{\gedSpeedupGCCkfive}{67.7}
\newcommand{\gedSpeedupGBCkHundred}{90.1}
\newcommand{\gedSpeedupGCCkHundred}{8.9}

\newcommand{\gedAlgoSpeedupGBC}{124.1}
\newcommand{\gedAlgoSpeedupGCC}{30.5}

\graphicspath{{plots/}}

\begin{document}

\title{\Large Group Centrality Maximization for Large-scale Graphs\thanks{
This work is partially supported by German Research Foundation (DFG) grant ME 3619/3-2
within Priority Programme 1736 \textit{Algorithms for Big Data} and by DFG grant
ME 3619/4-1 (\textit{Accelerating Matrix Computations for Mining Large Dynamic Complex Networks}).
}}
\date{}

\author{Eugenio Angriman\thanks{Humboldt-Universit\"at zu Berlin, Department of Computer Science, Germany.}
  \and Alexander van der Grinten\footnotemark[2]
  \and Aleksandar Bojchevski\thanks{Technical University of Munich, Department of Informatics, Germany.}
  \and Daniel Z\"ugner\footnotemark[3]
  \and Stephan G\"unnemann\footnotemark[3]
\and Henning Meyerhenke\footnotemark[2]}

\fancyfoot[R]{\scriptsize{Copyright \textcopyright\ 2020 by SIAM\\
Unauthorized reproduction of this article is prohibited}}

\maketitle

\begin{abstract}
The study of vertex centrality measures is a key aspect of network analysis.
Naturally, such centrality measures have been generalized to \emph{groups} of vertices;
for popular measures it was shown that the problem of finding the most central group
is $\mathcal{NP}$-hard.
As a result, approximation algorithms to maximize group centralities were introduced recently.
Despite a nearly-linear running time, approximation algorithms for group betweenness
and (to a lesser extent) group closeness are rather slow on large networks due to
high constant overheads.

That is why we introduce GED-Walk centrality,
a new submodular group centrality measure inspired by Katz centrality.
In contrast to closeness and betweenness, it considers walks of any
length rather than shortest paths,
with shorter walks having a higher contribution.
We define algorithms that (i) efficiently approximate the GED-Walk
score of a given group and (ii) efficiently approximate the
(proved to be $\mathcal{NP}$-hard) problem of finding a group with highest GED-Walk score.

Experiments on several real-world datasets show that scores obtained by GED-Walk improve performance on common graph mining tasks such as collective classification and graph-level classification.
An evaluation of empirical running times demonstrates that
maximizing GED-Walk (in approximation) is two orders of magnitude
faster compared to group betweenness approximation
and for group sizes $\leq 100$ one to two orders faster than group closeness approximation.
For graphs with tens of millions of
edges, approximate GED-Walk maximization typically needs less than one minute.
Furthermore, our experiments suggest that the maximization algorithms scale
linearly with the size of the input graph and the size of the group.
\\[0.25ex]
\end{abstract}

\section{Introduction}
\label{sec:intro}

Large network\footnote{We use the terms \enquote{graph}
	and \enquote{network} interchangeably in this paper.} data sets have become abundant in real-world applications;
consider online social networks, web and co-authorship graphs,
transportation and biological networks~\cite{newman2018networks},
to name just a few.
Network analysis techniques have become very important in the last one or two decades to gain insights
from these data~\cite{newman2018networks}.
One particularly interesting class of analysis methods are centrality measures:
they assign each vertex a numerical score that indicates the
importance of the vertex based on its position in the network.
A few very popular centrality measures are betweenness, closeness,
degree, eigenvector, PageRank, and Katz centrality~\cite{DBLP:journals/im/BoldiV14,newman2018networks}.

The notion of centrality has subsequently been extended to groups of
vertices~\cite{doi:10.1080/0022250X.1999.9990219};
this extension allows to determine the importance of whole
groups, not only of individuals and, vice versa, to determine important groups.
The latter question leads to an optimization problem: find the group of vertices
of a given size $k$ that maximizes a certain group centrality measure.
It is easy to imagine applications for group centrality,
\eg facility location and influence maximization.

\noindent \textbf{Motivation.}
Such optimization problems are often $\mathcal{NP}$-hard;
for example, this has been proven
for group betweenness~\cite{dolev2009incremental},
group closeness~\cite{chen2016efficient}
and even group degree.
Since exact results may not be necessary in real-world applications,
approximation is a reasonable way of dealing with complexity.
For example, in the case of group betweenness,
Mahmoody \etal~\cite{mahmoody2016scalable} designed a nearly-linear time
(in $|V|$ and $|E|$)
approximation algorithm based on random sampling, which has been shown to outperform
its competitors, while having similar accuracy.
Yet, due to constant factors, it is not applicable to really large real-world instances;
also (and in particular), the running time depends quadratically on $\log (1/\epsilon)$.
Indeed, the largest instances considered in their paper have around half a million edges.
While group \emph{closeness} approximation can already be done faster~\cite{bergamini2018scaling},
in the context of big data there is still a need for a meaningful group centrality measure with a
truly scalable algorithm for its optimization.

\noindent \textbf{Contribution.}
In this paper, we introduce a new centrality measure called ED-Walk for
\emph{exponentially decaying walk} (see Section~\ref{sub:group-cen-measures}).
While being inspired by Katz centrality, it allows a more natural extension
to the group case, then called GED-Walk.
GED-Walk takes walks of any length into account, but considers shorter ones more important.
We develop an algorithm to approximate the GED score of a given group
in nearly-linear time.
As our measure is submodular (see Section~\ref{sub:properties}),
we can design a greedy algorithm
to approximate a group with maximum GED-Walk score (in Section~\ref{sec:algorithm}).
To allow the greedy algorithm to quickly find the top-1 vertex with highest marginal gain,
we adapt a top-k centrality approach.
In our experiments, our algorithms are two orders of magnitude faster than the state of the art for
group betweenness approximation~\cite{mahmoody2016scalable}.
Compared to group closeness approximation~\cite{bergamini2018scaling},
we achieve a speedup of one to two orders of magnitude for
groups with at most 100 vertices.
Furthermore, our experiments indicate a linear running time behavior in practice (despite a
higher worst-case time complexity) for greedy GED-Walk approximation.
Finally, we show some potential applications of GED-Walk:
(i) choosing a training set with high group centrality leads to improved performance on the semi-supervised vertex classification task; and (ii) features derived based on GED-Walk improve graph classification performance.

\section{Problem Definition and Properties}
\label{sec:problem}

In this paper, we deal with unweighted graphs $G = (V, E)$; they can be directed or undirected.
By $A$, we denote the adjacency matrix of $G$.

\subsection{Group Centrality Measures.}
\label{sub:group-cen-measures}
We start by reviewing existing centrality measures and by
defining the new ED-Walk;
our main focus is then on the $\gedfn$ group centrality
derived from ED-Walk.

Closeness and betweenness, two of the most popular centrality measures,
are both based on shortest paths.
More precisely, betweenness is defined as the fraction of
shortest paths that cross a certain vertex (or a certain group of vertices, for
group betweenness).
Hence, betweenness can be computed using all-pairs shortest-path (APSP) techniques
and, interestingly, a sub-cubic algorithm for betweenness (on general graphs) would also
imply a faster algorithm for APSP~\cite{DBLP:conf/soda/AbboudGW15}.
Furthermore, the latter result holds true even if one considers the problem of computing
the betweenness centrality of only a single vertex of the graph.

Closeness, on the other hand, is defined as the reciprocal of the average distance
from a given vertex (or a group of vertices, for group closeness) to all other
vertices of the graph. Hence, computing the closeness of a single
vertex (or a single group of vertices) is easy; nonetheless,
unless the strong exponential time hypothesis (SETH) fails, there
is no sub-quadratic algorithm to find a vertex with maximal
closeness~\cite{borassi2016subquadratic}.

To avoid those complexity-theoretic barriers,
instead of deriving our measure from betweenness or closeness,
we will derive it from
Katz centrality (KC). KC is another popular centrality
measure but, in contrast to measures based on shortest paths,
it can be computed much more efficiently.
Let $\omega_i(x) = \sum_{z \in V} (A^i)_{xz}$ be the number of walks of length $i$
that start at vertex $x \in V$ (in our notation, we omit the
dependence of $\omega_i(x)$ on $G$; obviously, $\omega_i(x)$
potentially depends on the entire graph).\footnote{Katz centrality is
	sometimes alternatively defined based on
	walks \emph{ending} at a given vertex; from an algorithmic perspective,
	however, this difference is irrelevant.}
For a given parameter $\alpha \in \mathbb{R}$
(which needs to be chosen small enough~\cite{katz1953index}, see also
Proposition~\ref{prop:katz-relation}),
the Katz centrality of $x$ is defined as:
\begin{equation}
	\label{eq:katz-def}
	\katzfn(x) = \sum_{i=1}^\infty \alpha^i \sum_{z \in V} \left(A^i\right)_{xz}
			= \sum_{i=1}^\infty \alpha^i \omega_i(x).
\end{equation}
Note that $\katzfn$ does not only consider \emph{shortest paths}
but walks of any length.
Nevertheless, due to the fact that the contribution
of a walk decreases exponentially with its length,
short walks are still preferred.
As $\katzfn$ does not rely on APSP, it can be computed faster than $\bcfn$.
For example,
following from the algebraic formulation~\cite{katz1953index},
iterative linear solvers (\eg conjugate gradient)
can be used to compute $\katzfn$ for all vertices at the same time.
Furthermore, efficient nearly-linear time Katz approximation and ranking algorithms
exist~\cite{DBLP:conf/esa/GrintenBGBM18}. In fact, the algorithms that we present
in this paper will be based on those Katz algorithms.

In contrast to $\bcfn$, however, replacing the single vertex $x$ in
Eq.~(\ref{eq:katz-def}) by a group leads to a rather uninteresting measure:
as $\omega_i(x)$ and $\omega_i(x')$ count distinct walks for $x \neq x'$,
such a \enquote{group Katz} measure would just be equal to the sum of individual Katz
scores; a group with maximal score would just consist of the vertices
appearing in a top-$k$ Katz ranking.
Indeed, this contradicts one's intuition of group centralities: parts of a graph
that are \enquote{covered} by one vertex of the group should \emph{not} need
to be covered by a second group vertex.
Fortunately, we can construct a centrality satisfying this intuition
by replacing $\katzfn$ by a natural variant of it: instead of considering
walks that \emph{start} at a given vertex $x$, we consider all walks
that \emph{cross} vertex $x$. This leads to our definition of ED-Walk:
\begin{Definition}
\label{def:ed-walk}
	Let $\phi_i(S)$ be the number of $i$-walks (\ie walks of length $i$) that contain
	at least one vertex of $S \subseteq V$.
	Then, the ED-Walk (for \emph{exponentially decaying} walk) centrality
	for $x \in V$ is
	\[
	\edfn(x) := \sum_{i = 1}^{\infty}{\alpha^i}{\phi_{i}(\{x\})},
	\]
	where $\alpha > 0$ is a parameter of the centrality.
\end{Definition}

In the preceding definition, $\alpha$ needs to be chosen appropriately
small so that the series converges. In Section~\ref{sub:properties} we will
see that it is indeed possible to choose such an $\alpha$.
As claimed above, ED-Walk naturally generalizes to a group
measure:
\begin{Definition}[GED-Walk]
  \label{def:ged-walk}
	The \emph{GED-Walk centrality} of a group $S \subseteq V$ of vertices is given by:
	\begin{equation}
		\label{eq:ged_def}
		\gedfn(S) = \sum_{i = 1}^\infty \alpha^{i} \phi_{i}(S).
	\end{equation}
\end{Definition}

Two problems are specifically interesting in the context of group centralities:
(i) the problem of computing the centrality of a \emph{given} group $S \subseteq V$
and (ii) the problem of \emph{finding} a group that maximizes the group centrality measure.
We deal with both in this paper and in particular with the following optimization problem:
\begin{Definition}[GED-Walk maximization]
\label{def:ged-max}
Given a graph $G = (V, E)$ and an integer $k \geq 1$, find
\begin{equation}
\label{eq:ged-max}
  S^* = \argmax_{S \subseteq V, \vert S \vert = k} \gedfn(S).
\end{equation}
\end{Definition}

Note again that Problem~(\ref{eq:ged-max}) is in general different from finding the
$k$ vertices whose \emph{individual} ED-Walk centrality is highest (top-$k$ problem).
In fact, for closeness centrality, it was shown empirically that the top-$k$ problem and the
group closeness problem often have very different solutions~\cite{bergamini2018scaling}.
One may expect a similar behavior here as well
since a group $S$ with high score for GED-Walk is unlikely to have many members close to each other.
Such a ``close'' group would share many walks, leading to a relatively small $\phi_i(S)$.

\subsection{Properties of GED-Walk.}
\label{sub:properties}
Since GED is based on similar concepts as Katz centrality,
it is not surprising that the two measures are closely related.
Indeed, the convergence properties of $\katzfn(x)$ and GED-Walk
are identical:
\begin{proposition}
\label{prop:katz-relation}
	The following holds:
	\begin{enumerate}
		\item $\gedfn(V) = \sum_{x \in V} \katzfn(x)$
		\item If $\alpha < 1/\sigma_\mathrm{max}$, $\gedfn(S)$ is finite for all $S \subseteq V$,
			where $\sigma_\mathrm{max}$ is the largest singular value of the adjacency matrix of $G$.
	\end{enumerate}
\end{proposition}
Note that the equality of the first claim of Proposition~\ref{prop:katz-relation} does not hold
for groups smaller than $V$.
\begin{proof}
	To prove the first claim, it is enough to show that the walks
	that contribute to $\phi_i(V)$ and the walks that contribute
	to $\sum_{x \in V} \omega_i(V)$ are exactly the walks
	of length $i$ in $G$. Indeed, each walk $P = (x_0, x_1, x_2, \ldots, x_i)$
	of length $i$ contributes to $\phi_i(V)$. It also contributes to exactly
	one of the $\omega_i(x)$ for $x \in V$,
	namely to $\omega_i(x_0)$.
	After applying the first claim (and observing that $\gedfn(V)$ is an upper bound, also see Proposition~\ref{prop:ged_submod}),
	the second one follows from the well-known fact
	that $\katzfn(x)$ is finite for all $x \in V$
	iff $\alpha < 1/\sigma_\mathrm{max}$~\cite{katz1953index}.\hfill
\end{proof}

Now that we have established that GED-Walk is indeed well-defined,
let us prove some basic properties (defined in Appendix~\ref{apx:technical}) of the $\gedfn$ function:

\begin{proposition}
	\label{prop:ged_submod}
	GED-Walk is both non-decreasing and submodular as a set function.
\end{proposition}
\begin{proof}
	Monotonicity is obvious from Eq.~(\ref{eq:ged_def}): when the input set $S$ becomes larger,
	$GED(S)$ cannot decrease.
	To see that GED-Walk is also submodular, consider the marginal gain
	$\gedfn(S \cup \{x\}) - \gedfn(S)$, \ie the increase of the $\gedfn$
	score if a vertex $x$
	is added to the set $S$.
	This marginal gain is exactly the sum of the number of walks that contain
	$x$ but no vertex in $S$, with each walk weighted by a power of $\alpha$.
	As such, the marginal gain can only decrease if $S$ is replaced
	by a superset $T \supseteq S$.\hfill
\end{proof}

Given an algorithm to compute $\gedfn$, Proposition~\ref{prop:ged_submod}
would immediately allow us to construct a greedy algorithm to approximate
Problem~(\ref{eq:ged-max}) (using a well-known theorem on
submodular approximation~\cite{nemhauser1978analysis},
also see Proposition~\ref{prop:greedy_approx} in Appendix~\ref{apx:technical}).
Nevertheless, as we will see in Section~\ref{sub:max-ged},
we can do better than naively applying the greedy approach.
Before discussing any algorithms, however, we will first demonstrate that
maximizing GED-Walk to optimality is $\mathcal{NP}$-hard; hence,
approximation is indeed appropriate to solve Problem~(\ref{eq:ged-max}).
\begin{theorem}
\label{thm:hardness}
Solving Problem~(\ref{eq:ged-max}) to optimality is $\mathcal{NP}$-hard.
\end{theorem}
Intuitively, Theorem~\ref{thm:hardness} holds because if $\alpha$ is chosen
small enough, GED-Walk degrades to group degree
(or, equivalently, vertex cover).\footnote{A set of vertices
has a group degree of $|E|$ iff it is a vertex cover.}
\begin{proof}
	Let $G$ be a graph.
	Let $\alpha < \frac1{|V|^3+|V|}$.
	We show that $G$ has a vertex cover of size $k$
	if and only if there is a group $S$ of $k$ vertices with
	$\gedfn(S) \geq \alpha |E|$.

	First, assume that $G$ has a vertex cover $S$ of size $k$.
	To see now that a group with GED $\geq \alpha \vert E \vert$ indeed exists,
	consider $S$: since $S$ covers all edges, the contribution of the term corresponding to $i=1$
	in Eq.~(\ref{eq:ged_def}) is already at least $\alpha \vert E \vert$ (independent of $\alpha$).

	\sloppy
	Secondly, assume that
	a group $S$ with $\gedfn(S) \geq \alpha |E|$ exists.
	Let $E(S)$ denote the set of edges incident to at least one vertex in $S$.
	$\gedfn(S)$ is given by $\alpha |E(S)| + \sum_{i=2}^\infty \alpha^i \phi_i(S)$.
	Thus, to decide if $S$ is a vertex cover, it is sufficient to show that $\sum_{i=2}^\infty \alpha^i \phi_i(S) < \alpha$: in this case, $|E(S)|$
	needs to be $|E|$ as all walks of length $\geq 2$
	together cannot match the contribution of any $1$-walk to $\gedfn(S)$.
	It holds that $\sum_{i=2}^\infty \alpha^i \phi_i(S)
		\leq \sum_{i=2}^\infty \alpha^i |V|^{i+1}
		= (|V| \sum_{i=0}^\infty \alpha^i |V|^i) - |V|(1 + \alpha |V|)
		= |V|\frac1{1 - \alpha|V|} - |V|(1 + \alpha |V|)$.
	A straightforward calculation shows that the latter term is smaller than
	$\alpha$ if $\alpha < \frac1{|V|^3+|V|}$.\hfill
\end{proof}

\section{Algorithms for GED-Walk}
\label{sec:algorithm}

\subsection{Computing GED-Walk Centrality.}
\label{sub:compute-ged}
\newcommand{\hitphi}{\phi^{\mathrm{hit}}}%
\newcommand{\missphi}{\phi^{\mathrm{miss}}}%
\newcommand{\misspsi}{\psi^{\mathrm{miss}}}%
In this section, we discuss the problem of computing $\gedfn(S)$ for a given
group $S$.
As we are not aware of an obvious translation of our $\gedfn(S)$ definition
to a closed-form expression, we employ an approximation algorithm
that computes the infinite series $\gedfn(S)$ up to an arbitrarily
small additive error $\epsilon > 0$.
To see how the algorithm works, let $\ell \in \mathbb{N}$ be a positive integer.
We split the
series $\gedfn(S)$ into the first $\ell$ terms and a tail consisting of
an infinite number of terms. This allows us to reduce
the problem of approximating $\gedfn(S)$
to the problem of (exactly) computing the $\ell$-th partial sum
$\gedfn_{\leq \ell}(S) := \sum_{i=1}^\ell \alpha^i \phi_i(S)$
and to the problem of finding an upper bound on the tail
$\gedfn_{> \ell}(S) := \sum_{i=\ell+1}^\infty \alpha^i \phi_i(S)$.
In particular, we are looking for an upper bound that converges to zero
for increasing $\ell$.
Given such an upper bound, our approximation algorithm chooses $\ell$ such that this bound
is below the specified error threshold $\epsilon$. The
algorithm returns $\gedfn_{\leq \ell}(S)$ once such an $\ell$ is found.

\subsubsection{Computation of $\phi_i(S)$.}
\label{sub:compute-phi}
In order to compute $\gedfn_{\leq \ell}(S)$, it is obviously enough to compute
$\phi_i(S)$ for $i \in \{1, \ldots, \ell\}$.
The key insight to computing $\phi_i(S)$ efficiently is that it can
be expressed as a series of recurrences (similar to Katz centrality):
let us define $\hitphi_i(x, S)$ as the number of $i$-walks
ending in $x$ and containing at least one vertex from $S$. Obviously, the walks
that contribute to $\hitphi_i$ partition the walks that contribute
to $\phi_i$ in the sense that
\begin{equation}
	\label{eq:phisum}
	\phi_i(S) = \sum_{x \in V} \hitphi_i(x, S).
\end{equation}
On the other hand, $\hitphi$ can be expressed in terms of $\missphi_i(x, S)$,
\ie the number of $i$-walks that end in $x$ but do not contain
any vertex from $S$:
\[
  \hitphi_{i}(v, S) =
  \begin{cases}
    \sum_{(u, v) \in E}{\hitphi_{i-1}(u, S) + \missphi_{i-1}(u, S)} & v \in S\\
    \sum_{(u, v) \in E}{\hitphi_{i-1}(u, S)} & v \notin S
  \end{cases}
\]

This takes advantage of the fact that any $i$-walk is the concatenation
of a $(i-1)$-walk and a single edge. The expression considers $(i-1)$-walks which do not
contain a vertex in $S$ only if the concatenated edge ends in $S$.
Finally, $\missphi_i(x, S)$ can similarly
be expressed as the recurrence:
\[
  \missphi_{i}(v, S) =
  \begin{cases}
    0 & v \in S\\
    \sum_{(u, v) \in E}{\missphi_{i-1}(u, S)} & v \notin S
  \end{cases}
\]
In the $v \in S$ case, all walks ending in $v$ have a vertex in $S$, so
no walks contribute to $\missphi$. We remark that the $i = 1$ base cases
of $\hitphi$ and $\missphi$ can easily be computed directly from $G$. Collectively,
this yields an $\mathcal{O}(\ell(|V|+|E|))$-time algorithm for computing the
GED-Walk score of a given group.
We also note that the computation of $\hitphi_i$ and $\missphi_i$
can be trivially parallelized by computing the values of those
functions for multiple vertices in parallel.

\subsubsection{Bounding $\gedfn_{> \ell}(S)$.}
\label{sub:bounds}
\newcommand{\bound}[1]{\widehat{#1}}
To complete our algorithm, we need a sequence of upper bounds on
$\gedfn_{> \ell}(S)$ such that the upper bound converges to zero for
$\ell \to \infty$.
Using Proposition~\ref{prop:katz-relation}, we can lift tail bounds of Katz
centrality to GED-Walk. In particular, applying the first claim of this proposition
shows:
\begin{equation}
\label{eq:katz-bound}
\gedfn_{> \ell}(S) \leq \gedfn_{> \ell}(V)
	= \sum_{x \in V} \sum_{i=\ell+1}^\infty \alpha^i \omega_i(x).
\end{equation}
Bounds for the latter term can be found in the literature on Katz centrality.
For example, in~\cite{DBLP:conf/esa/GrintenBGBM18}, it was shown that
$\sum_{i=\ell+1} \omega_i(x)
	\leq \frac{\deg_\mathrm{max}}{1 - \alpha \deg_\mathrm{max}}
			\alpha^{\ell+1} \omega_\ell(x)$,
for $\alpha < 1/\deg_\mathrm{max}$. If we apply Eq.~(\ref{eq:katz-bound}),
this yields the bound
\begin{equation}
\label{eq:comb-bound}
\gedfn_{> \ell}(V)
	\leq \alpha^{\ell+1} \frac{\deg_\mathrm{max}}{1 - \alpha\ \deg_\mathrm{max}}
			\sum_{x \in V} \omega_\ell(x)
\end{equation}
for GED-Walk; we call this bound
the \emph{combinatorial bound}
(due to the nature of the proof in~\cite{DBLP:conf/esa/GrintenBGBM18}).
Here, $\deg_\mathrm{max}$ denotes the maximum degree of any
vertex in $G$.
We generalize this
statement to arbitrary $\alpha$
(\ie $\alpha < 1/\sigma_\mathrm{max}$), at the cost of a factor of
$\sqrt{|V|}$.
For completeness, we show the proof in the case of $\gedfn$.

\begin{lemma}
\label{lem:spectral-bound}
It holds that:
\begin{equation}
\label{eq:spectral-bound}
\gedfn_{> \ell}(V)
	\leq \sqrt{|V|} \alpha^{\ell+1} \frac{\sigma_\mathrm{max}}{1 - \alpha\ \sigma_\mathrm{max}}
			\sum_{x \in V} \omega_\ell(x).
\end{equation}
\end{lemma}
We call the bound of Eq.~(\ref{eq:spectral-bound}) the \emph{spectral bound}.
\begin{proof}
First note that $\sum_{x \in V} \omega_i(x) = |A^i \mathbf{1}_{|V|}|_1$,
where $\mathbf{1}_{|V|}$ denotes the vector containing all ones in $\mathbb{R}^{|V|}$
and $|\,.\,|_1$ denotes the 1-norm.
Furthermore, let $||\,.\,||_p$ denote the induced $p$-norm on $\mathbb{R}$-valued
matrices. Since $||\,.\,||_p$ is compatible with $|\,.\,|_p$, it holds that
$\sum_{x \in V} \omega_{\ell + i}(x) \leq ||A^i||_1 |A^\ell \mathbf{1}_{|V|}|_1
	= ||A^i||_1 \sum_{x \in V} \omega_\ell(x)
	\leq \sqrt{|V|} ||A||_2^i \sum_{x \in V} \omega_\ell(x).$
Here, the last inequality uses the fact that $||M||_1 \leq \sqrt{n} ||M||_2$
for $n \times n$ matrices $M$~\cite{feng2003norms}
and the fact that $||\,.\,||_2$ is sub-multiplicative.
Furthermore, it is well-known that $||A||_2 = \sigma_\mathrm{max}$;
hence
\[
	\sum_{x \in V} \omega_{\ell + i}(x)
			\leq \sqrt{|V|} \sigma_\mathrm{max} \sum_{x \in V} \omega_\ell(x)
\]
Substituting this into Eq.~(\ref{eq:katz-bound}), we get:
$\gedfn_{>\ell}(S) \leq \sum_{i = 1}^\infty \alpha^{\ell + i} \sum_{x \in V} \omega_{\ell + i}(x)
	\leq \sqrt{|V|} \sum_{i = 1}^\infty
		\alpha^{\ell + i} \sigma_\mathrm{max}^i \sum_{x \in V} \omega_\ell(x)$.
After rewriting the geometric series
$\sum_{i = 1}^\infty (\alpha\ \sigma_\mathrm{max})^i
	= \alpha\ \sigma_\mathrm{max} / (1 - \alpha\ \sigma_\mathrm{max})$,
we obtain the statement of the lemma.\hfill
\end{proof}

\subsubsection{Complexity Analysis.}
\begin{algorithm}[t]
\small
\caption{GED-Walk Computation}
\label{algo:compute-ged}
\textbf{Input:} Graph $G = (V, E)$, parameters $\alpha$, $\epsilon$ and group $S \subseteq V$\\
\textbf{Output:} $\gedfn(S) \pm \epsilon$

\begin{algorithmic}[1]
	\State $c \gets 0$
	\State $i \gets 1$
	\Loop
		\State \label{line:comp-phi} compute $\phi_i(S)$
		\State $c \gets c + \alpha^i \phi_i(S)$
		\State \label{line:comp-bound} compute bound
			$B_\ell$ s.t. $\gedfn_{> \ell}(S) \leq B_\ell$
		\If{$B_\ell < \epsilon$}
			\State \Return $c$
		\EndIf
		\State $i \gets i + 1$
	\EndLoop
\end{algorithmic}
\end{algorithm}

Algorithm~\ref{algo:compute-ged} shows the pseudocode of the algorithm.
Line~\ref{line:comp-phi} computes $\phi_i(S)$ using Equation~(\ref{eq:phisum})
and the recurrences for $\missphi_i$ and $\hitphi_i$ from Section~\ref{sub:compute-phi}.
Assuming that $\missphi_{i-1}$ and $\hitphi_{i-1}$ are stored,
this can be done in time $\mathcal{O}(|V| + |E|)$.
Line~\ref{line:comp-bound} computes an upper bound on $\gedfn_{> \ell}(S)$;
we use either the combinatorial or the spectral bound from Section~\ref{sub:bounds}.

\begin{lemma}
	\label{lem:running-time}
	The $\gedfn$ score of a given group can be approximated
	up to an additive error of $\epsilon$
	in
	$\mathcal{O}(\frac{\log (|V|/\epsilon)}{\log 1/(\alpha \sigma_\mathrm{max})})$
	iterations.
\end{lemma}
\begin{proof}
	We assume that the bound of Lemma~\ref{lem:spectral-bound} is used in the algorithm
	(similar results can be obtained for other bounds).
	By construction, the algorithm terminates once $\gedfn_{> \ell}(S)
		\leq \sqrt{|V|} \alpha^{\ell+1} \frac{\sigma_\mathrm{max}}{1 - \alpha\ \sigma_\mathrm{max}}
				\sum_{x \in V} \omega_\ell(x)$.
	Note that $\omega_\ell(x) \leq \sqrt{|V|} (\sigma_\mathrm{max})^\ell$; thus, we
	can bound the entire $\gedfn_{> \ell}(S)$ term by
	$|V|^{2.5} (\alpha \sigma_\mathrm{max})^{\ell+1}
		\frac{\sigma_\mathrm{max}}{1 - \alpha\ \sigma_\mathrm{max}}$.
	A straightforward calculation shows that for
	$\ell > 2.5 \frac{\log (|V|/\epsilon)}{\log 1/{\alpha \sigma_\mathrm{max}}}$,
	the previous term becomes smaller than $\epsilon$.\hfill
\end{proof}
\begin{corollary}
	The $\gedfn$ score of a given group can be approximated
	up to an additive error of $\epsilon$
	in time
	$\mathcal{O}(\frac{\log (|V|/\epsilon)}{\log 1/(\alpha \sigma_\mathrm{max})}(|V| + |E|))$.
\end{corollary}

\subsection{Maximizing GED-Walk Centrality.}
\label{sub:max-ged}
\newcommand{\gedscore}{\mathtt{c}}
\newcommand{\margscore}{\mathtt{c'}}
\newcommand{\exactvector}{\mathtt{r}}

Let $\gedfn(S, x)$ denote the marginal gain of a vertex $x$
\wrt a group $S$; in other words, $\gedfn(S, x) := \gedfn(S \cup \{x\}) - \gedfn(S)$.
As observed in Section~\ref{sub:properties}, the properties given by
Proposition~\ref{prop:ged_submod} imply that
a greedy algorithm that successively picks a vertex $x \in V$ with highest
marginal gain $\gedfn(S, x)$
yields a $(1 - \frac 1e)$-approximation
for Problem~(\ref{eq:ged-max}).
Note, however, that we do not have an algorithm
to compute $\gedfn(S, x)$ exactly. While we could use the approximation
algorithm of Section~\ref{sub:compute-ged} to feed approximate
values of $\gedfn(S, x)$ into a greedy algorithm,
this procedure would involve more work than necessary: indeed, the
greedy algorithm does not care about the value of $\gedfn(S, x)$ at all,
it just needs to ascertain that there is no $x' \in V$ with $\gedfn(S, x') > \gedfn(S, x)$.

This consideration is similar to a top-1 centrality
problem.\footnote{Top-1 algorithms have been developed for multiple centrality
measures, \eg
in~\cite{bergamini2019computing}.}
Hence, we adapt the ideas of the top-$k$ Katz ranking algorithm
that was introduced
in~\cite{DBLP:conf/esa/GrintenBGBM18} to the case of $\gedfn(S, x)$.\footnote{Note that
	the vertex that maximizes $\gedfn(S, x) = \gedfn(S \cup \{x\}) - \gedfn(S)$
	is exactly the vertex that maximizes $\gedfn(S \cup \{x\})$. However,
	algorithmically, it is advantageous to deal with $\gedfn(S \cup \{x\})$, as it
	allows us to construct a lazy greedy algorithm (see Section~\ref{sub:lazy-greedy}).}
Applied to the marginal gain of GED-Walk, the main ingredients of
this algorithm are families
$L_\ell(S, x)$ and $U_\ell(S, x)$ of lower and upper bounds on $\gedfn(S, x)$,
satisfying the following definition:
\begin{Definition}
\label{defn:suitable-bounds}
	We say that families of functions $L_\ell(S, x)$ and $U_\ell(S, x)$
	are \emph{suitable} bounds on the marginal gain of GED-Walk if
	the following conditions are all satisfied:
	\begin{enumerate}
		\item[(i)] $L_\ell(S, x) \leq \gedfn(S, x) \leq U_\ell(S, x)$
		\item[(ii)] $\lim_{\ell \to \infty} L_\ell(S, x)
			= \lim_{\ell \to \infty} U_\ell(S, x) = \gedfn(S, x)$
		\item[(iii)] $L_\ell(S, x)$ and $U_\ell(S, x)$ are non-increasing in $S$
	\end{enumerate}
\end{Definition}
In addition to $L_\ell$ and $U_\ell$, we need the following definition:
\begin{Definition}
\label{defn:separation}
	Let $\epsilon > 0$ and fix some $\ell \in \mathbb{N}$.
	Let $x, x' \in V$ be two vertices. If
	$L_\ell(S, x) > U_\ell(S, x') - \epsilon$, we say that $x$ is $\epsilon$-separated
	from $x'$.
\end{Definition}

Given the definition, it is easy to see that if there is a vertex $x \in V$
that is $\epsilon$-separated from all other $x' \in V$, then either $x$
is the vertex with top-1 marginal gain $\gedfn(S, x)$ or the
marginal gain of the top-1 vertex
is at most $\gedfn(S, x) + \epsilon$ (in fact, this follows directly
from the first property of Definition~\ref{defn:suitable-bounds}).
Note that the introduction of $\epsilon$ is required here to
guarantee that separation can be achieved for finite $\ell$
even if $x$ and $x'$ have truly identical marginal gains.
Furthermore, while the first condition of Definition~\ref{defn:suitable-bounds}
is required for $\epsilon$-separation to work, the second and third
conditions are required for the correctness of the algorithm that
we construct in Section~\ref{sub:lazy-greedy}.
\subsubsection{Construction of $L_\ell(S, x)$ and $U_\ell(S, x)$.}
\label{sub:lu-bounds}
In order to apply the methodology from Section~\ref{sub:max-ged}, we need
suitable families $L_\ell(S, x)$ and $U_\ell(S, x)$ of bounds.
Luckily, it turns out that we can re-use the bounds developed in
Section~\ref{sub:bounds}.
Indeed, it holds that:
$\gedfn(S, x)
	= \gedfn_{\leq \ell}(S \cup \{x\}) - \gedfn_{\leq \ell}(S)
		+ \gedfn_{> \ell}(S \cup \{x\}) - \gedfn_{> \ell} (S)
	\geq \gedfn_{\leq \ell}(S \cup \{x\}) - \gedfn_{\leq \ell}(S)$.
In the above calculation,
$\gedfn_{> \ell}(S \cup \{x\}) - \gedfn_{> \ell}(S) \geq 0$
because $\phi_i$ is non-decreasing as a set function.
Hence
$L_\ell(S, x) := \gedfn_{\leq \ell}(S \cup \{x\}) - \gedfn_{\leq \ell}(S)$
yields a family of lower bounds on the marginal gain of $\gedfn$.
On the other hand,
$\gedfn(S, x)
	= \gedfn_{\leq \ell}(S \cup \{x\}) - \gedfn_{\leq \ell}(S)
		+ \gedfn_{> \ell}(S \cup \{x\}) - \gedfn_{> \ell} (S)
	\leq \gedfn_{\leq \ell}(S \cup \{x\}) - \gedfn_{\leq \ell}(S)
		+ \gedfn_{> \ell}(S \cup \{x\})$.
Thus, a family of upper bounds on the marginal gain of $\gedfn$ is given by
$U_\ell(S, x) := \gedfn_{\leq \ell}(S \cup \{x\}) - \gedfn_{\leq \ell}(S) + B_\ell(V)$,
where $B_\ell(V)$ denotes either the combinatorial or
the spectral bound developed in Section~\ref{sub:max-ged}.

\begin{lemma}
	Let $\gedfn_{\leq \ell}(S, x) := \gedfn_{\leq \ell}(S \cup \{x\}) - \gedfn_{\leq \ell}(S)$.
	The two families
	\begin{align*}
		L_\ell(S, x) &:= \gedfn_{\leq \ell}(S, x) \\
		U_\ell(S, x) &:= \gedfn_{\leq \ell}(S, x) + B_\ell(V)
	\end{align*}
	form suitable families of bounds on the marginal gain of $\gedfn$.
\end{lemma}
\begin{proof}
	The discussion at the beginning of this section already
	implies that properties (i) and (ii) of Definition~\ref{defn:suitable-bounds}
	hold.
	For seeing that (iii) also holds, it is enough to show that
	$\gedfn_{\leq \ell}(S, x)$ is non-increasing as a set function; however,
	this directly follows from the proof of Proposition~\ref{prop:ged_submod}).\hfill
\end{proof}

\subsubsection{Lazy-Greedy Algorithm.}
\label{sub:lazy-greedy}
\begin{algorithm}[tb]
\small
\caption{Lazy greedy algorithm for GED-Walk Maximization}
\label{algo:lazy_greedy}
\textbf{Input:} Graph $G = (V, E)$, parameter $\alpha$, and group size $k$.\\
\textbf{Output:} \ Group $S \subseteq V$ with $\gedfn(S) \geq (1 - 1/e)$ times the optimum score.

\begin{algorithmic}[1]
	\newcommand{\true}{\mathbf{true}}
	\newcommand{\false}{\mathbf{false}}
	\newcommand{\kwNot}{\mathbf{not}}
	\newcommand{\pqExtract}{\mathtt{extractMax}}
	\newcommand{\pqInsert}{\mathtt{insert}}
	\State $S \gets \emptyset$
	\State $\ell \gets 1$
	\State $L(x) = \infty, U(x) \gets \infty$ for all $x \in V$
	\State $Q_L \gets V$, $Q_U \gets V$ \Comment{Priority queues \wrt $L$, $U$}
	\smallskip
	\Procedure{lazyUpdate}{$Q$} \label{line:lazy-update}
		\Repeat
			\State $x \gets Q.\mathbf{top}$
			\State $L(x) \gets L_\ell(S, x)$
			\State $U(x) \gets U_\ell(S, x)$
			\State $Q_L.\mathbf{update}(x)$; $Q_U.\mathbf{update}(x)$
		\Until{$x = Q.\mathbf{top}$}
		\State \Return $x$
	\EndProcedure
	\smallskip
	\Loop
		\State reset $L(x) \gets \infty$; $U(x) \gets \infty$ for all $x \in V$
			\label{line:lu-reset}
		\State $Q_L.\mathbf{rebuild}()$; $Q_U.\mathbf{rebuild}()$
		\Loop
			\If{$|S| = k$}
				\State \Return $S$
			\EndIf
			\State $u \gets \Call{lazyUpdate}{Q_L}$
			\State $Q_L.\mathbf{remove}(u)$; $Q_U.\mathbf{remove}(u)$
			\State $v \gets \Call{lazyUpdate}{Q_U}$
			\If{$L(u) \leq U(v) - \epsilon/k$} \label{line:separation-check}
				\Statex \Comment{$u$ is not $(\epsilon/k)$-separated from $v$; increase $\ell$}
				\State $Q_L.\mathbf{insert}(u)$; $Q_U.\mathbf{insert}(u)$
				\State \textbf{break}
			\EndIf
			\State $S \gets S \cup \{u\}$
		\EndLoop
		\State $\ell \gets 2 \ell$ \Comment{Increase $\ell$ using geometric progression}
			\label{line:ell-increment}
	\EndLoop
\end{algorithmic}
\end{algorithm}

Taking advantage of the ideas from Section~\ref{sub:lu-bounds}, we can construct a
greedy approximation algorithm for GED-Walk maximization.
To reduce the number of evaluations of our objectives
(without impacting the solution or the (worst-case) time complexity),
we use a lazy strategy inspired by the well-known lazy greedy
algorithm for submodular approximation~\cite{minoux1978accelerated}.
However, in contrast to the standard lazy greedy algorithm,
we do not evaluate our submodular objective function $\gedfn$
directly; instead, we apply Definition~\ref{defn:separation}
to find the top-1 vertex with highest marginal gain.
To this end, instead of lazily evaluating $\gedfn(S, x)$
for vertices $x$,
we lazily evaluate $L_\ell(S, x)$ and $U_\ell(S, x)$.
In fact, to apply Definition~\ref{defn:separation}, we need to find
the vertex $u \in V \setminus S$ that maximizes $L_\ell(u)$
and the vertex $v \in V \setminus S, v \neq u$
that maximizes $U_\ell(v)$.
Hence, we rank all vertices according to priorities $L(x) \geq L_\ell(S, x)$
and $U(x) \geq U_\ell(S, x)$ and lazily compute the true
values of $L_\ell(S, x)$ and $U_\ell(S, x)$ until
$u$ and $v$ are identified.

Algorithm~\ref{algo:lazy_greedy} depicts the pseudocode of this algorithm.
Procedure \textsc{lazyUpdate} (line~\ref{line:lazy-update}) lazily updates
the priorities $L(x)$
and $U(x)$ until the exact values of $L_\ell(S, x)$
and $U_\ell(S, x)$ are known for the topmost element of the given
priority queue.
In line~\ref{line:separation-check}, the resulting vertices $u, v$
are checked for $(\epsilon/k)$-separation. It is necessary
to achieve an $(\epsilon/k)$-separation (and not only an
$\epsilon$-separation) to guarantee that the total absolute
error is below $\epsilon$, even after all $k$ iterations.
If separation fails, $\ell$ is incremented in line~\ref{line:ell-increment}
and we reset $L(x)$ and $U(x)$ for all $x \in V$ in line~\ref{line:lu-reset};
otherwise, $u$ (\ie the vertex with top-1 marginal gain) is added to $S$.

\begin{lemma}
	If $L_\ell$ and $U_\ell$ are suitable families
	of bounds, Algorithm~\ref{algo:lazy_greedy} computes a
	group $S$ such that
	$\gedfn(S) \geq (1 - \frac 1e) \gedfn(S^*) - \epsilon$,
	where $S^*$ is the group with highest $\gedfn$ score.
\end{lemma}
\begin{proof}
	First, note that it is indeed sufficient to only consider
	the vertices $u, v$ with highest $L_\ell(S, u)$ and $U_\ell(S, v)$
	for $(\epsilon/k)$-separation: if those vertices are not
	$(\epsilon/k)$-separated, $u$ is not the top-1 vertex with highest marginal gain
	(up to $\epsilon/k$).
	Due to property (ii) of Definition~\ref{defn:suitable-bounds},
	$u$ and $v$ will eventually be $(\epsilon/k)$-separated and the
	algorithm terminates.
	Hence, it is enough to show that the algorithm indeed finds the
	vertices $u, v$ with highest $L_\ell(S, u)$ and $U_\ell(S, v)$.
	This follows from the invariant that $L(x) \geq L_\ell(S, x)$
	and $U(x) \geq U_\ell(S, x)$. The invariant obviously
	holds after each reset of $L(x)$ and $U(x)$. Furthermore,
	property (iii) of Definition~\ref{defn:suitable-bounds} guarantees
	that it also holds when the set $S$ is extended.
	The approximation guarantee follows from
	Proposition~\ref{prop:additive-greedy} in the appendix.
	Specifically, it follows from a variant of the classical
	result on submodular approximation, adapted to the
	case of an additive error in the objective function.\hfill
\end{proof}

\paragraph{Initialization of $L(x)$ and $U(x)$.}
To accelerate the algorithm in practice, it is crucial that the initial
values of $L(x)$ and $U(x)$ are chosen appropriately during each reset:
if we reset $L(x) = U(x) = \infty$ as in Algorithm~\ref{algo:lazy_greedy},
the algorithm has to evaluate
$L_\ell(S, x)$ and $U_\ell(S, x)$ for all $x \in V \setminus S$
whenever $\ell$ increases!
To avoid this issue, we use $\missphi_i(x, S)$
to provide a better initialization of $L(x)$ and $U(x)$.
Let $\misspsi_i(x, S)$ be the value of $\missphi_i(x, S)$ in the reverse
graph of $G$, \ie $\misspsi_i(x, S)$ is the number of $i$-walks
that \emph{start} at $x$ but do not contain any vertex of $S$.
The following lemma summarizes our initialization strategy:

\begin{lemma}
\label{lem:initialize}
	Let
	\[
		P_i(S, x) := \sum_{j = 0}^i \missphi_{i - j}(x, S)\ \misspsi_{j}(S, x).
	\]
	The following holds:
	\begin{enumerate}
		\item $P_i(S,x) \geq \phi_i(S \cup \{x\}) - \phi_i(S)$

		\item Let $B_\ell$ denote the bound from the construction of $U_\ell$.
			$P_i$ yields the following bounds on $L_\ell$ and $U_\ell$:
			\begin{align*}
				L_\ell(S, x) &\leq \sum_{i=1}^\ell \alpha^i P_i(S, x) \\
				U_\ell(S, x) &\leq \sum_{i=1}^\ell \alpha^i P_i(S, x) + B_\ell(V).
			\end{align*}
	\end{enumerate}
\end{lemma}
Thus, instead of initializing $L(x) = U(x) = \infty$, we compute
$P_i(S, x)$ and use the right-hand sides of the second statement
of Lemma~\ref{lem:initialize}
as initial values for $L(x)$ and $U(x)$.
Note that using the recurrence for $\missphi_i$ from Section~\ref{sub:compute-ged}
(and a similar recurrence for $\misspsi_i$), $P_\ell(S, x)$ can be computed in time
$\mathcal{O}(\ell (|V| + |E|))$, simultaneously for all $x$.
\begin{proof}
	To prove the first claim, observe that the difference between
	$P_i(S, x)$ and $\phi_i(S \cup \{x\}) - \phi_i(S)$
	is that $P_i(S, x)$ counts walks that contain
	cycles multiple times while $\phi_i(S \cup \{x\}) - \phi_i(S)$
	counts those walks only once. The second claim follows immediately.\hfill
\end{proof}

\paragraph{Complexity of the lazy algorithm.}
Assume that Algorithm~\ref{algo:lazy_greedy} uses
the spectral bound from Lemma~\ref{lem:spectral-bound}.
The worst-case value of $\ell$ to achieve $(\epsilon/k)$-separation
is given by Lemma~\ref{lem:running-time}, namely
$\mathcal{O}(\frac{\log (k |V|/\epsilon)}{\log 1/(\alpha \sigma_\mathrm{max})})$.
Note that the algorithm does some iterations that do not
achieve $(\epsilon/k)$-separation.
We calculate the cost of each iteration in terms of the number of evaluations
of $\phi_i$. In this measure, the cost of each iteration is $\ell$.
Thus, as $\ell$ is doubled on unsuccessful $(\epsilon/k)$-separation,
the cost of all unsuccessful attempts to achieve
$(\epsilon/k)$-separation
is always smaller than the cost of a
successful attempt to achieve $(\epsilon/k)$-separation.
Hence, the worst-case running time of
Algorithm~\ref{algo:lazy_greedy} is
$\mathcal{O}(k |V| \frac{\log (k |V|/\epsilon)}{\log 1/(\alpha \sigma_\mathrm{max})} (|V|+|E|))$:
in each of the $k$ successful iterations,
it can be necessary to extract all $\mathcal{O}(|V|)$
elements of the priority queues
and evaluate $L_\ell(x)$ and $U_\ell(x)$ on each of them.

We expect that much fewer evaluations of $L_\ell(x)$ and $U_\ell(x)$
will be required in practice than in the worst-case:
in fact, we expect almost all
vertices to have low marginal gains (\ie only few walks cross those vertices)
and $\gedfn$ will never be evaluated on those vertices.
Hence, we investigate the practical performance of Algorithm~\ref{algo:lazy_greedy}
in Section~\ref{sec:experiments}.

\section{Experiments}
\label{sec:experiments}
Our algorithms for GED-Walk maximization are implemented in C++
on top of the open-source framework NetworKit~\cite{staudt2016networkit},
which also includes implementations of the aforementioned algorithms for group betweenness
and group closeness maximization.
All experiments are executed on a Linux server with two Intel Xeon Gold 6154 CPUs (36 cores in total) and
1.5 TB of memory.
If not stated differently, every experiment uses 36 threads (one per core), and the default algorithm for
$\gedfn$-Walk maximization uses the combinatorial bound (see Section~\ref{sub:bounds}) and the lazy-greedy
strategy (see Section~\ref{sub:lazy-greedy}) with $\epsilon = 0.5$.
All networks are undirected; real-world networks have been downloaded from the Koblenz Network
Collection~\cite{DBLP:conf/www/Kunegis13}, and synthetic networks have been generated using the graph generators
provided in NetworKit (see Appendix~\ref{apx:experimental_details} for more details about the settings of the
parameters of the generators).
Because group closeness is not defined on disconnected graphs (without modifications), we always consider the largest connected component
of our input instances.

\subsection{Scalability \wrt Group Size.}
\label{sub:scalability_k}
\begin{figure}[t]
\begin{subfigure}[t]{\columnwidth}
\centering
\includegraphics{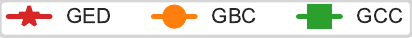}\par
\end{subfigure}
\begin{subfigure}[t]{.5\columnwidth}
\centering
\includegraphics{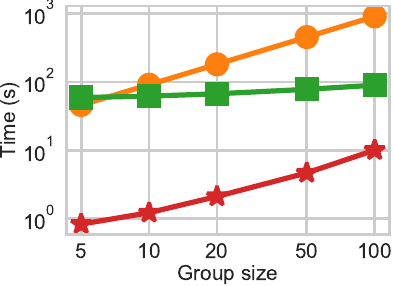}
\caption{Running time (s) of $\gedfn$-Walk, $\gccfn$, and $\gbcfn$ maximization (note that $k$ is in log-scale).}
\label{fig:group_size_scalability}
\end{subfigure}%
\hfill
\begin{subfigure}[t]{.5\columnwidth}
\centering
\includegraphics{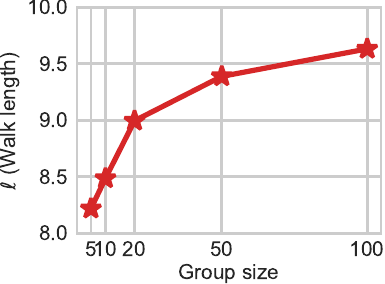}
\caption{Length of the walks considered by our algorithm for $\gedfn$-Walk maximization.}
\label{fig:nlevels}
\end{subfigure}
\caption{Scalability \wrt $k$ of $\gedfn$-Walk, $\gbcfn$, and $\gccfn$ maximization
(Figure~\ref{fig:group_size_scalability}), and highest walk length considered by our GED-Walk
maximization algorithm (Figure~\ref{fig:nlevels}).
Data points are aggregated using the geometric mean over the instance in Table~\ref{tab:rt_instances_table}.
}
\end{figure}

Figure~\ref{fig:group_size_scalability} shows the average running time in seconds of $\gbcfn$, $\gedfn$-Walk,
and $\gccfn$ maximization for group sizes from $5$ to $100$.
For small group sizes $\gedfn$-Walk can be maximized much faster than $\gbcfn$ and $\gccfn$: for $k=5$
our algorithm for $\gedfn$-Walk maximization is on average
\speed{\gedSpeedupGBCkfive} faster than $\gbcfn$ maximization, and
\speed{\gedSpeedupGCCkfive} faster than $\gccfn$ maximization, while for $k = 100$ it is on average
\speed{\gedSpeedupGBCkHundred} faster than $\gbcfn$ maximization and \speed{\gedSpeedupGCCkHundred}
faster than $\gccfn$ maximization.
On the other hand, $\gccfn$ maximization scales better than $\gbcfn$ and $\gedfn$-Walk maximization
\wrt the group size (see Appendix~\ref{apx:stochastic-greedy} for additional results on large groups).
This behavior is expected since the evaluation of marginal gains of $\gccfn$
becomes computationally cheaper for larger groups.
This does not apply to our algorithm for maximizing $\gedfn$-Walk, which instead needs to increase
the length of the walks $\ell$ while the group grows
(see Algorithm~\ref{algo:lazy_greedy}, line~\ref{line:ell-increment}).

Yet, one can also observe that the group closeness score increases only very slowly (and more slowly
than $\gedfn$-Walk's) when increasing the group size (see Figure~\ref{fig:scores} in the appendix).
This means that, from a certain group size on, the choice of a new group member hardly makes a difference
for group closeness -- in most cases only the distance to very close vertices can be reduced.
In that sense $\gedfn$-Walk seems to distinguish better between more and less promising candidates.
Furthermore, Figure~\ref{fig:nlevels} shows the length of the walks $\ell$ considered by our algorithm \wrt $k$.
In accordance with what we stated in Section~\ref{sub:lazy-greedy}, $\ell$ grows sub-linearly with $k$.

\subsection{Scalability to Large (Synthetic) Graphs.}
\label{sub:graph-scalability}
\begin{figure}[t]
\centering
\begin{subfigure}[t]{.5\columnwidth}
\includegraphics{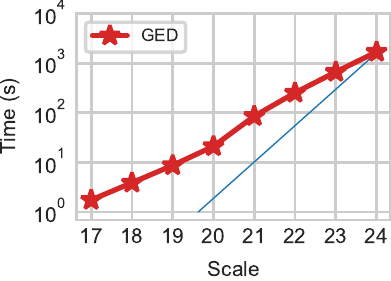}
\caption{Running time (s) on \erdosr networks.}
\label{fig:run_time_er}
\end{subfigure}%
\hfill
\begin{subfigure}[t]{.5\columnwidth}
\includegraphics{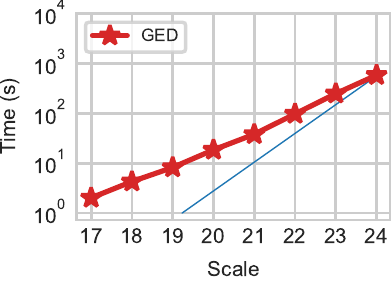}
\caption{Running time (s) on R-MAT networks.}
\label{fig:run_time_rmat}
\end{subfigure}

\begin{subfigure}[t]{.5\columnwidth}
\includegraphics{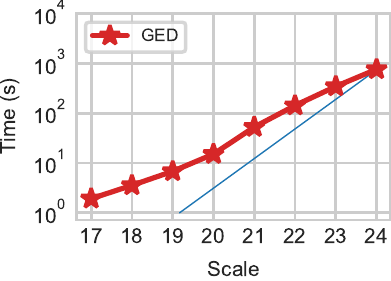}
\caption{Running time (s) on \ba networks.}
\label{fig:run_time_ba}
\end{subfigure}%
\hfill
\begin{subfigure}[t]{.5\columnwidth}
\includegraphics{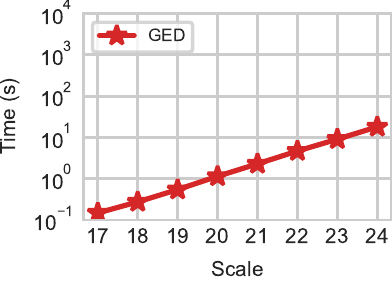}
\caption{Running time (s) on random hyperbolic networks.}
\label{fig:run_time_hyp}
\end{subfigure}
\caption{Running time in seconds on 36 cores of $\gedfn$-Walk maximization on synthetic networks
with $2^{17}$ to $2^{24}$ vertices, $k = 10$.
Data points are aggregated using the geometric mean over three different randomly generated networks
(see Appendix~\ref{apx:experimental_details} for further details).}
\label{fig:run_time_synt}
\end{figure}

Figure~\ref{fig:run_time_synt} shows the running time in seconds of $\gedfn$-Walk maximization with
$k = 10$ on randomly generated networks
using the \erdosr, R-MAT~\cite{chakrabarti2004r} and \ba~\cite{barabasi1999emergence} models
as well as the random hyperbolic generator from von Looz
\etal~\cite{von2016generating}.
The thin blue lines represent the linear regression on the running times (all resulting
$p$-values are  $< 10^{-3}$ and therefore small
enough for statistical relevance); \wrt the running time curves,
the regression lines either have a steeper
slope (Figures~\ref{fig:run_time_er},~\ref{fig:run_time_rmat}, and~\ref{fig:run_time_ba}),
or they match it almost perfectly in the case of the hyperbolic random generator.
Therefore, for the network types and sizes under consideration,
GED-Walk maximization scales (empirically) linearly \wrt $|V|$.

Experimental results on large real-world networks with hundreds of million of edges can be found
in Table~\ref{tab:large_inst_table} in Appendix~\ref{apx:additional_exp}. Our algorithm needs up to
six minutes to finish, in most cases only around one minute.

\subsection{Parallel Scalability}

\begin{figure}[t]
\centering
\begin{subfigure}[t]{\columnwidth}
\centering
\includegraphics{runtime_legend}\par
\end{subfigure}
\begin{subfigure}[t]{.5\columnwidth}
\includegraphics{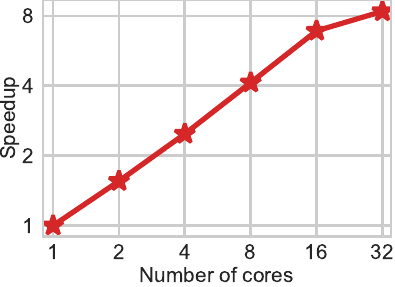}
\caption{Multi-core speedups of $\gedfn$ maximization over single-core
$\gedfn$ maximization.}
\label{fig:ged_multi_core_scalability}
\end{subfigure}%
\hfill
\begin{subfigure}[t]{.5\columnwidth}
\includegraphics{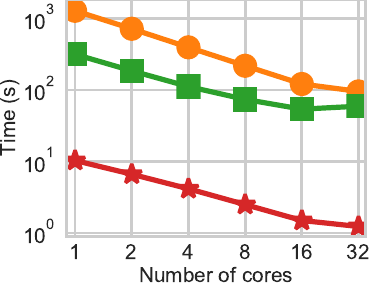}
\caption{Running time of $\gbcfn$, $\gedfn$, and $\gccfn$ maximization
\wrt the number of cores.}
\label{fig:multi_core_scalability}
\end{subfigure}
\caption{Multi-core scalability of $\gbcfn$, $\gedfn$, and $\gccfn$ maximization, $k = 10$.
Data points are aggregated using the geometric mean over the instances in Table~\ref{tab:rt_instances_table}.}
\end{figure}

As stated above, the GED-Walk algorithm can be parallelized by
computing $\hitphi_i(v)$ and $\missphi_i(v)$ in parallel
for different $v \in V$.
Figure~\ref{fig:ged_multi_core_scalability} shows the parallel speedup of our algorithm for
maximizing $\gedfn$-Walk over itself running on
a single core for $k = 10$.
The scalability is moderate up to 16 cores, while on 32 cores it does not gain much additional speedup.
A similar behavior was observed before for group closeness~\cite{bergamini2018scaling}.
Figure~\ref{fig:multi_core_scalability} shows the running time in seconds of $\gbcfn$, $\gedfn$, and
$\gccfn$ with increasing number of cores, for $k=10$.
On a single core, $\gedfn$-Walk maximization is on average \speed{\gedAlgoSpeedupGBC} faster than $\gbcfn$ maximization
and \speed{\gedAlgoSpeedupGCC} faster than $\gccfn$ maximization.
As we increment the number of cores,
the decrement of the running time of the three algorithms is comparable -- except $\gccfn$ on 32 cores:
In this case, GCC is slower than with 16 cores on the considered instances.
The limited scalability affecting all three algorithms
is probably due to memory latency becoming a bottleneck in the execution on multiple
cores~\cite{bader2005architectural, lumsdaine2007challenges}.
Figure~\ref{fig:multi_core_scalability} also shows that our algorithm for $\gedfn$-Walk maximization
finishes on average within a few seconds on $16$ cores.
Here the running time of the algorithm is dominated by its sequential parts, and it is not surprising
that adding $16$ more cores does not speed the algorithm up substantially.

\subsection{Applications of GED-Walk.}
In this section we demonstrate the usefulness of GED-Walk for different applications by showing that it improves the performance on two important graph mining tasks: semi-supervised vertex classification and graph classification. As a preprocessing step, for both tasks before applying GED-Walk we first construct a weighted graph using the symmetrically normalized adjacency matrix $D^{-1/2}A{D^{-1/2}}$ which is often used in the literature \cite{zhou2004learning,kipf2016semi}, where $D$ is a diagonal matrix of vertex degrees.
Here, instead of taking the contribution of a $i$-walk $(e_1, \ldots, e_i)$ to be $\alpha$,
we define it to be $\alpha \prod_{j = 1}^i w(e_j)$ where $w(e_j) \leq 1$ denotes the
weight of edge $e_j$. Except for the introduction of coefficients
in the recurrences of $\hitphi$ and $\missphi$, no modifications
to our algorithm are required.
Compared to our unweighted definition of GED-Walk,
this weighted variant converges even faster, as the contribution of each walk is smaller.

\subsubsection{Vertex classification.}
Semi-supervised vertex classification is a fundamental graph mining problem where the goal is to predict the class labels of all vertices in a graph given a small set of labelled vertices and the graph structure.
The choice of which vertices we label (\ie which vertices we include in the training set) before building a classification model can have a significant impact on the test accuracy, especially when the number of labelled vertices is small in comparison to the size of the graph \cite{avrachenkov2013choice, shchur2018pitfalls}.
Since many models rely on diffusion to propagate information on the graph, our hypothesis is that selecting a training set with high group centrality will improve diffusion, and thus also improve accuracy.
To test this hypothesis we evaluate the performance of Label Propagation \cite{zhou2004learning, chapelle2009semi} given different strategies for choosing the training set. We choose the simple Label Propagation model as our benchmark to isolate the effect on accuracy more clearly,  but similar conclusions apply to more sophisticated vertex classification models such as graph neural networks \cite{kipf2016semi}.

More specifically, we evaluate the classification accuracy on two common benchmark graphs: Cora ($|V|=2810, |E|=7981$) and Wiki ($|V|=2357, |E|=11592$) \cite{sen2008collective}. We use the Normalized Laplacian variant of Label Propagation \cite{zhou2004learning} setting the value for the return probability hyper-parameter to $0.85$. We let the vertices with highest group centrality according to GED-Walk be in the training set and the rest of the vertices be in the test set. We compare GED-Walk with the following baselines for selecting the training set: \textit{RND}: select vertices at random (results averaged over 10 trials); \textit{DEG}: select the vertices with highest degree; \textit{NBC}: select vertices with highest (non-group) betweenness centrality; \textit{GBC}: select vertices with highest group betweenness centrality; and \textit{PPR}: select vertices with highest PageRank.

\begin{figure}[b]
	\centering
	\begin{subfigure}[t]{.5\columnwidth}
		\includegraphics[width=\columnwidth]{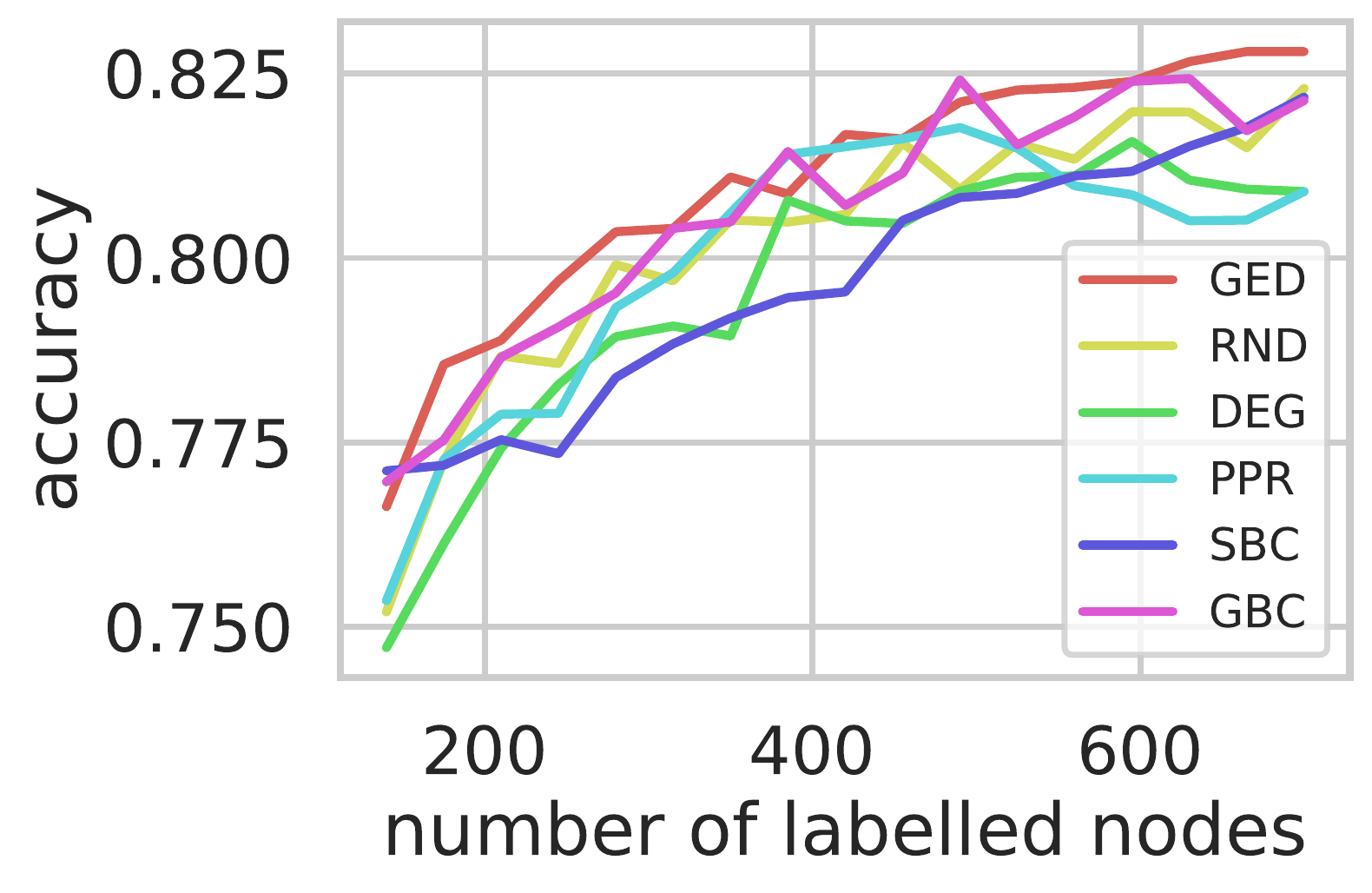}
		\caption{Cora}
		\label{fig:node_classification_cora}
	\end{subfigure}%
	\hfill
	\begin{subfigure}[t]{.5\columnwidth}
		\includegraphics[width=\columnwidth]{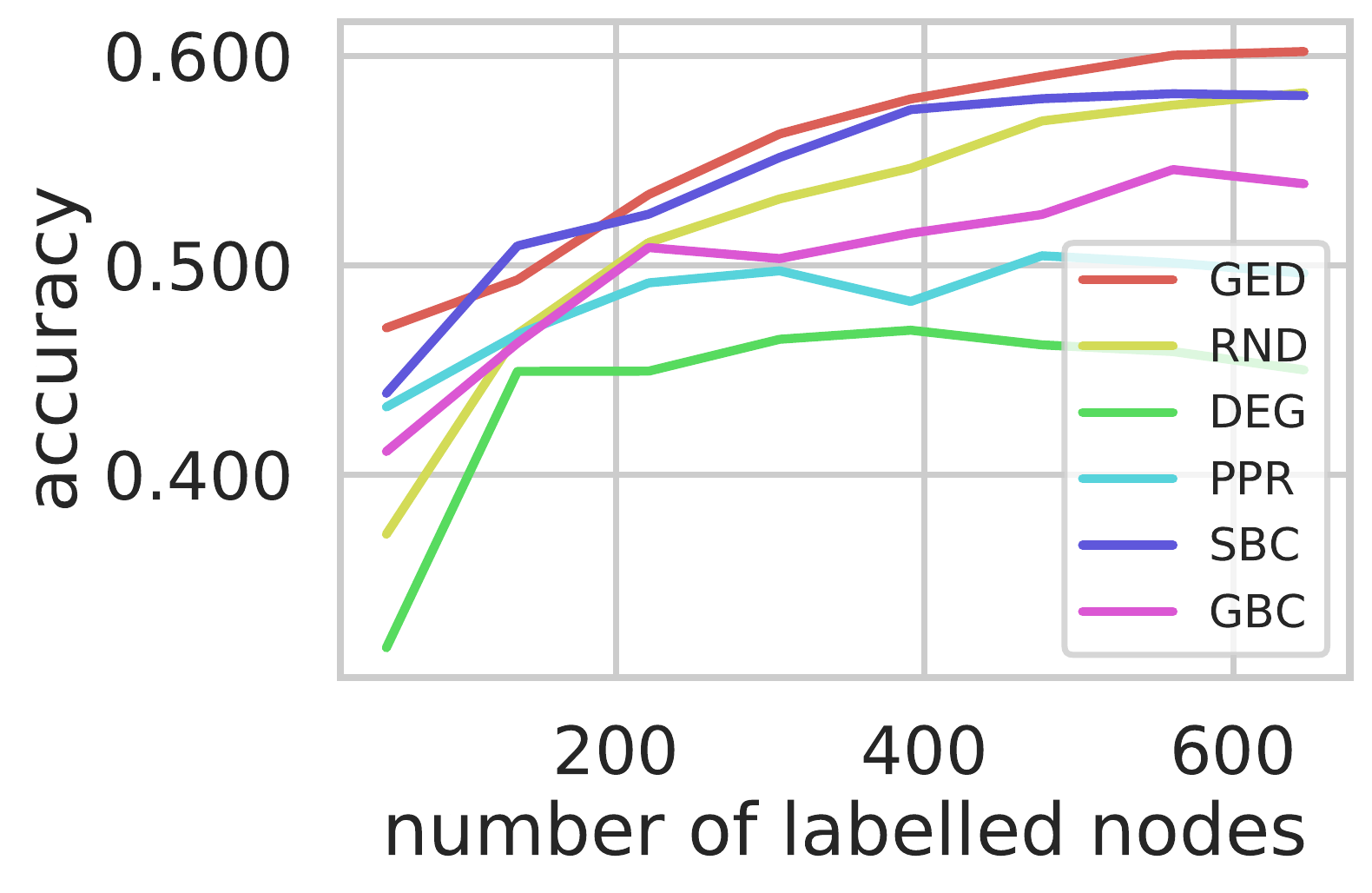}
		\caption{Wiki}
		\label{fig:node_classification_wiki}
	\end{subfigure}
	\caption{Semi-supervised vertex classification accuracy for different strategies for choosing the training set.}
	\label{fig:node_classification}
\end{figure}

On Figure~\ref{fig:node_classification}, we can see that for both datasets and across different number of labelled vertices, selecting the training set using GED-Walk leads to highest (or comparable) test accuracy. Furthermore, while
the second-best baseline strategy is different on the different datasets (on Cora it is the \textit{GBC} strategy and on Wiki its the \textit{SBC} strategy), GED-Walk is consistently better. Overall, these results confirm our hypothesis.

\subsubsection{Graph classification.}
Graph classification is another fundamental graph mining problem where the goal is to classify entire graphs based on features derived from their topology/structure.
In contrast to the vertex classification task, where each dataset is a single graph, here each dataset consists of many graphs of varying size and their associated ground-truth class labels.
Our hypothesis is that the vertices with high group centrality identified by GED-Walk capture rich information about the graph structure and can be used to derive features that are useful for graph classification.

To extract features based on GED-Walk, we first obtain the group of $k$ vertices with highest (approximate) group centrality score.
The group centrality score itself is the first feature we extract.
In addition we summarize the marginal gains of all remaining vertices in the graph in a histogram with $b$ bins.
We concatenate these features to get a feature vector $\mathbf{x}_i \in \mathbb{R}^{b+1}$ for each graph $i$ in the dataset.
These features are useful since graphs with similar structure will have similar group scores and marginal gains.
We denote this base case by \textsc{GED}.

In addition, we obtain the topic-sensitive PageRank vector of each graph, where we specify the teleport set to be equal to the vertices in the group with highest (approximate) group centrality.
Then we summarize this vector using (a) a histogram of $b$ bins and (b) by extracting the top $p$ values, denoted by \textsc{PPR}-H and \textsc{PPR}-T, respectively. Intuitively, these features capture the amount of diffusion on the graph.

As a strong baseline we compute the eigenvalues of the adjacency matrix and summarize them (a) in a histogram of $b$ bins and (b) by extracting the top $p$ eigenvalues,
denoted by \textsc{Eig}-H and \textsc{Eig}-T, respectively.
This is inspired by recent work on graph classification \cite{gallandinvariant} showing that spectral features can outperform deep-learning based approaches.
The ability to efficiently compute the eigenvalue histograms further motivates using them as a feature for graph classification \cite{network_density}. Similarly, a strong advantage of the features based on GED-Walk is that we can efficiently compute them.
Last, we also combine the spectral- and GED-based features by concatenation, i.e.\ \textsc{Eig}-T+\textsc{GED} denotes the combination of \textsc{Eig}-T and \textsc{GED} features.

In the following experiments we fix the value of the hyper-parameters $k=10$, $b=20$, and $p=10$; however,
in practice these parameters can also be tuned e.g.\ using cross-validation.
We split the data into 80\% training and 20\% test set and average the results for 10 independent random splits.

Table~\ref{tab:graph_classification} summarizes the graph classification results.
We observe that enhancing the baseline features with our GED-Walk-based features improves the classification performance on all datasets,
with variants using all available features, i.e.\  \textsc{Eig}+\textsc{GED}+\textsc{PPR} performing best.
Moreover, as shown in Table~\ref{tab:all_graph_classification} in the appendix, using the most central group as determined by GED-Walk as teleport set yields performance improvements over standard PageRank with teleport to all vertices.
See Table~\ref{tab:graph_classification_datasets} in the appendix for an overview of datasets used.
In summary, we have shown that GED-Walk captures meaningful information about the graph structure that is complementary to baseline spectral features.
We argue that GED-Walk can be used as a relatively inexpensive to compute additional source of information to enhance any existing graph classification model.

\begin {table}
\caption{Graph classification accuracy (in \%).
Best performance per dataset marked in bold.}
\label{tab:graph_classification}
\centering
\footnotesize
\begin{tabular}{lrrrrr}
    \toprule
    Dataset &  ENZ. &  IMD. &  Mut. &  PRO. &  RED. \\
    \midrule
    \textsc{Eig}-T                           &    23.02 &        56.59 &         56.90 &     73.35 &          75.31 \\
    \textsc{Eig}-H                           &    23.47 &        70.28 &         68.88 &     72.42 &          72.02 \\ \midrule
    \textsc{Ged}                             &    19.18 &        60.64 &         64.51 &     71.95 &          70.59 \\
    \textsc{Ged+PPR}$^*$-H                   &    20.85 &        65.18 &         65.46 &     72.18 &          71.59 \\
    \textsc{Ged+PPR}$^{**}$-H                &    20.39 &        66.27 &         65.86 &     72.44 &          75.95 \\\midrule
    \textsc{Eig}-T+\textsc{Ged}-T            &    26.46 &        63.56 &         64.14 &\textbf{74.06} &      80.18 \\
    \textsc{Eig}-H+\textsc{Ged}-H            &    23.14 &        69.74 &         69.08 &     73.13 &          75.16 \\\midrule
    \textsc{Eig}-T+ & \multirow{2}{*}{27.45} & \multirow{2}{*}{69.25} & \multirow{2}{*}{62.78} &\multirow{2}{*}{73.54} &\multirow{2}{*}{76.70}\\
    \textsc{Ged+PPR}$^*$-T\\
    \addlinespace[3pt]
    \textsc{Eig}-H+    &\multirow{2}{*}{24.12} &\multirow{2}{*}{\textbf{71.62}}&\multirow{2}{*}{\textbf{69.18}}& \multirow{2}{*}{73.10} &\multirow{2}{*}{74.17} \\
    \textsc{Ged+PPR}$^*$-H\\
    \addlinespace[3pt]
    \textsc{Eig}-T+ &\multirow{2}{*}{\textbf{27.88}} &\multirow{2}{*}{68.53} &\multirow{2}{*}{62.43} &\multirow{2}{*}{73.72}&\multirow{2}{*}{80.48} \\
    \textsc{Ged+PPR}$^{**}$-T\\
    \addlinespace[3pt]
    \textsc{Eig}-H+ &\multirow{2}{*}{24.78}&\multirow{2}{*}{70.54}&\multirow{2}{*}{68.81}&\multirow{2}{*}{72.97}&\multirow{2}{*}{\textbf{81.43}}\\
    \textsc{Ged+PPR}$^{**}$-H\\
    \bottomrule
    \multicolumn{6}{l}{$^*$: PageRank teleport probability 0.85;}\\
    \multicolumn{6}{l}{$^{**}$: PageRank teleport probability 0.15}
\end{tabular}
\end{table}

\section{Related Work}
\label{sec:relatedwork}

Several algorithms to maximize group centralities are found in the literature,
and many of them are based on greedy approaches.
Chehreghani \etal~\cite{chehreghani2018depth} provide an extensive analysis of several
algorithms for group betweenness centrality estimation, and present a new algorithm based
on an alternative definition of distance between a vertex and a group of vertices.
Greedy approximation algorithms to maximize group centrality also
exist for measures like closeness (by Bergamini \etal~\cite{bergamini2018scaling})
and current-flow closeness
(by Li \etal~\cite{li2019current}).
Puzis \etal~\cite{puzis2007finding} state an algorithm for group betweenness
maximization that does not
utilize submodular approximation but relies on a branch-and-bound approach.

Alternative group centrality measures were introduced by Ishakian \etal~\cite{ishakian2012framework},
Puzis \etal~\cite{puzis2007fast}, and Fushimi \etal~\cite{fushimi2018new}.
Ishakian \etal defined single-vertex centrality measures based on a generic concept of \emph{path},
and generalized them to groups of vertices.
However, in contrast to GED-Walk, those measures are defined for directed acyclic graphs only.
Puzis \etal defined the \emph{path betweenness} centrality of a group of vertices
by counting the fraction
of shortest paths that cross all vertices within the group.
The proposed algorithms to find a group with
high path-betweenness are quadratic in the best case (both in time and memory)
and therefore cannot scale to large-scale networks.
In fact, the largest graphs considered in their paper have only 500 vertices.
Fushimi \etal proposed a new measure called \emph{connectedness centrality},
that targets the specific problem of
identifying the appropriate locations where to place evacuation facilities on road networks.
For this reason, their measure assumes the existence of edge failure probabilities and is
not applicable to general graphs. Due to expensive Monte Carlo simulation,
computing their measure
also requires many hours on graphs with 100000 edges,
while GED-Walk scales to hundreds of millions of edges.

Network design is a different strategy to address the group centrality maximization problem when the input
graph is allowed to be altered.
Medya \etal~\cite{medya2018group} introduced two algorithms that optimize the increment of the
coverage centrality of a given group of vertices by iteratively adding $k$ new edges to the input graph.

In the context of single-vertex centrality, multiple algebraic-based metrics that do not
only depend on shortest paths exist. Besides the aforementioned Katz,
some examples are \emph{current-flow closeness}~\cite{DBLP:conf/stacs/BrandesF05}, \emph{PageRank},
\emph{flow betweenness}~\cite{freeman1991flowbc}
and \emph{random-walk betweenness}~\cite{newman2005randombc}.

\section{Conclusions}
\label{sec:conclusions}
We have presented a new group centrality measure, GED-Walk, and
efficient approximation algorithms for its optimization and for computing
the $\gedfn$ value of a given group.
GED-Walk's descriptive power is demonstrated by experiments on two fundamental graph mining tasks:
both semi-supervised vertex classification and graph classification benefit from the new measure.
As GED can be optimized faster than earlier measures, it is often a viable
replacement for more expensive measures in performance-sensitive applications.

More precisely, in terms of running time, our algorithm for GED-Walk maximization
significantly outperforms the state-of-the-art algorithms for maximizing group closeness and group
betweenness centrality when group sizes are at most $100$.
The fact that $\gedfn$ scales worse than $\gccfn$ \wrt to $k$ may seem as a limitation;
however, we expect that many applications are interested in group sizes
considerably smaller than $100$.

Experiments on synthetic networks indicate that our algorithm for GED-Walk maximization
scales linearly with the number of vertices. For graphs with $2^{24}$ vertices
and more than $100$M edges, it needs up to half an hour -- often less.
In fact, our algorithm can maximize GED-Walk for small groups on real-world graphs with hundreds
of millions of edges within a few minutes.
In future work we plan to apply GED-Walk for network and traffic monitoring \cite{dolev2009incremental, puzis2013augmented};
for developing immunization strategies to lower the vulnerability of a network to epidemic outbreaks \cite{pastor2002immunization}; and
for improving landmark-based shortest path queries \cite{goldberg2005computing}.

\bibliographystyle{acm}
\bibliography{references}

\clearpage

\appendix

\section{Technical Details}
\label{apx:technical}
The following definition is well-known in the literature:
\begin{Definition}[submodularity, monotonicity]
Let $X$ be a non-empty finite set.
\begin{itemize}
  \item A set function $f: 2^X \to \mathbb{R}$ is called \emph{submodular} if for every
    $S,T \subseteq X$ and every $x \in X \setminus T$ we have that
    $f(S \cup \{x\}) - f(S) \geq f(T \cup \{x\}) - f(T)$.
    In this context, the value $f(T \cup \{x\}) - f(T)$ is called the
    \emph{marginal gain} of $x$ (\wrt the set $T$).

  \item A set function $f: 2^X \to \mathbb{R}$ is called \emph{non-decreasing} if for
    every $S \subseteq T$ we have that $f(S) \leq f(T)$.
	Similarly, $f$ is called \emph{non-increasing} if for every $S \subseteq T$,
	we have $f(S) \geq f(T)$.
\end{itemize}
\end{Definition}

The significance of this definition stems from the fact that there is a classical result about optimizing non-decreasing submodular functions:
\begin{proposition}
	\label{prop:greedy_approx}
	\cite{nemhauser1978analysis}
	Let $f: 2^X \to \mathbb{R}$ be a non-decreasing submodular set function. Consider the
	problem of maximizing $f$ over all subsets $S \subseteq X$
	\wrt the cardinality constraint $|S| \leq k$ for some
	$k \in \mathbb{N}$. Let $S^*$ be the optimal
	solution to this problem, in other words:
	\[S^* = \argmax_{S \subseteq X, |S| \leq k} f(S)\]
	Then the greedy algorithm that constructs
	$S$ by iteratively adding the element $x \in X$ with highest
	marginal gain $f(S, x) := f(S \cup \{x\}) - f(S)$ to $S$, yields a
	$(1-1/e)$-approximation for this problem. Specifically,
	if $\tilde{S}$ is the result of the greedy algorithm,
	it holds that $f(\tilde{S}) \geq (1-1/e) f(S^*)$.
\end{proposition}

Proposition~\ref{prop:greedy_approx} can be shown using the following characterization
of submodular functions:
\begin{lemma}
	\label{lem:submod_char}
	\cite{nemhauser1978analysis}
	Let $f$ be a set function. $f$ is non-decreasing and submodular
	if and only if for all $T, S$:
	$f(T) \leq f(S) + \sum_{x \in T \setminus S} f(S, x)$.
\end{lemma}

As we do not have access to the exact values of $f$ in the case of $\gedfn$,
we need the following variant of the classical result:
\begin{proposition}
	\label{prop:additive-greedy}
	Let $\epsilon > 0$.
	Consider the greedy algorithm of Proposition~\ref{prop:greedy_approx}
	but assume that instead of picking an element $x$ that maximizes
	$f$, it only picks an element $x'$
	such that $f(S, x') \geq f(S, x) - \epsilon/k$.
	Let $\tilde S$ be the set returned by this algorithm. It holds that
	$f(\tilde{S}) \geq (1-1/e) f(S^*) - \epsilon$.
\end{proposition}
In the case of a relative error, this proposition is
proved in~\cite{goundan2007revisiting} (Theorem 1);
the proof follows~\cite{nemhauser1978analysis}.
Our proof is mostly a mechanical translation to the case of an additive error.
\begin{proof}
	Let $\tilde S_i$ be the set that the algorithm constructs in iteration $i$.
	Denote by $\rho_i$ the marginal gain that the algorithm achieves in step $i$,
	\ie $\rho_i = f(\tilde S_i) - f(\tilde S_{i-1})$. Furthermore, let
	$\rho^*_i = \max_{x \in X} f(\tilde S_{i-1}, x)$, \ie $\rho^*_i$ is the marginal
	gain that would be achieved if an element with truly maximal
	marginal gain was known in step $i$. Due to the construction of the algorithm, it
	holds that $\rho_i \leq \rho^*_i \leq \rho_i + \epsilon/k$.

	Apply Lemma~\ref{lem:submod_char} to the sets $S^*$ and $\tilde S_j$. This yields
	\begin{align*}
		f(S^*) &\leq f(\tilde S_j) + \sum_{x \in S^* \setminus \tilde S_j} f(\tilde S_j, x) \\
			&= \sum_{i=1}^j \rho_i + \sum_{x \in S^* \setminus \tilde S_j} f(\tilde S_j, x).
	\end{align*}
	As the second sum in this term sums over at most $k$ elements
	that are all smaller than $\rho^*_{j+1}$, it holds that:
	\[ f(S^*) \leq \sum_{i=1}^j \rho_i + k\left(\rho_{j+1} + \frac\epsilon k\right)
		\leq \sum_{i=1}^j \rho_i + k \rho_{j+1} + \epsilon, \]
	or, equivalently:
	\[ \rho_{j+1} \geq \frac 1k f(S^*) - \frac 1k \sum_{i=1}^j \rho_i - \frac \epsilon k. \]
	Adding $\sum_{i=1}^j \rho_i$ on both sides yields:
	\[
		\sum_{i=1}^{j+1} \rho_i
			\geq \frac 1k f(S^*) + \frac {k-1}k \sum_{i=1}^j \rho_i - \frac \epsilon k.
	\]
	Using this equation, it is straightforward to prove the following claim by induction:
	\[ \sum_{i=1}^j \rho_i \geq \frac{k^j - (k-1)^j}{k^j} f(S^*) - j \frac \epsilon k. \]
	Finally, substituting $j$ with $k$, using
	$\left(\frac{k-1}k\right)^k \leq 1/e$, and rewriting $\sum_{i=1}^k \rho_i = f(\tilde S)$
	proves the proposition.\hfill
\end{proof}

\section{Details of the Experimental Setup}
\label{apx:experimental_details}

\begin{table}[tb]
\setlength{\tabcolsep}{5pt}
\footnotesize
\caption{Real-world instances used for our experiments.
Sizes correspond to the largest connected component of every instance.}
\label{tab:rt_instances_table}
\centering
\begin{tabular}{lrrr}
\toprule
Network & $|V|$ & $|E|$ & Category\\
\hline
dimacs9-COL & \numprint{435666} & \numprint{521200} & Road\\
munmun\_twitter\_social & \numprint{465017} & \numprint{833540} & Social\\
com-dblp & \numprint{317080} & \numprint{1049866} & Co-author\\
wikipedia\_link\_mr & \numprint{92875} & \numprint{1396893} & Hyperlink\\
roadNet-PA & \numprint{1087562} & \numprint{1541514} & Road\\
citeseer & \numprint{365154} & \numprint{1721981} & Citation\\
roadNet-TX & \numprint{1351137} & \numprint{1879201} & Road\\
web-Stanford & \numprint{255265} & \numprint{1941926} & Hyperlink\\
petster-dog-household & \numprint{255968} & \numprint{2148090} & Social\\
wikipedia\_link\_bn & \numprint{225970} & \numprint{2183246} & Hyperlink\\
petster-catdog-household & \numprint{324249} & \numprint{2642635} & Social\\
wikipedia\_link\_uz & \numprint{439263} & \numprint{2920885} & Hyperlink\\
\toprule
\end{tabular}
\end{table}

\begin{table}[tb]
    \footnotesize
    \centering
	\caption{Graph classification datasets.}
	\begin{tabular}{lrr}
		Dataset 							& \# Graphs & \# Classes\\ \toprule
		Mutagenicity \cite{mutag1,mutag2}	& 4,337		& 2 		\\
		PROTEINS \cite{proteins1,proteins2}	& 1,113		& 2			\\
		ENZYMES	\cite{proteins2,enzymes}	& 600		& 6			\\
		IMDB-BINARY \cite{Yanardag_DGK}		& 1,000		& 2			\\
		REDDIT-BINARY \cite{Yanardag_DGK} 	& 2,000		& 2			\\ \bottomrule
	\end{tabular}
	\label{tab:graph_classification_datasets}
\end{table}

Tables~\ref{tab:rt_instances_table} and~\ref{tab:graph_classification_datasets}
state details about our real-world instances.
In the following, we describe the settings that were used to generate
the synthetic graphs for the experiment of Section~\ref{sub:graph-scalability}:
for every number of vertices, we generate three random networks by setting three different random seeds.
For the \erdosr generator, we set as probability parameter $p = 20 / |V|$.
For the R-MAT generator, the parameter setting is the same
as in the Graph 500's benchmark~\cite{murphy2010introducing} (\ie edge factor 16,
$a = \numprint{0.57}$, $b = \numprint{0.19}$, $c = \numprint{0.19}$,
and $d = \numprint{0.05}$).
For the \ba generator, we set the average degree to $20$, and for the random hyperbolic generator
we set the average degree to $20$, and the exponent of the power-law distribution to 3.

\section{Dealing with Large $k$}
\label{apx:stochastic-greedy}

\begin{figure}[t]
\centering
\begin{subfigure}[t]{\columnwidth}
\centering
\includegraphics{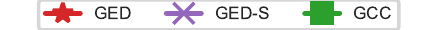}
\end{subfigure}
\begin{subfigure}[t]{.5\columnwidth}
\includegraphics{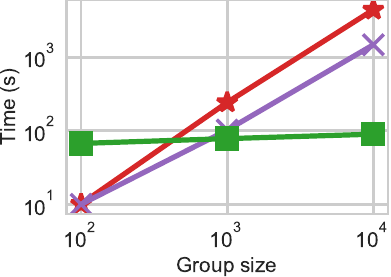}
\caption{Running time (s) of $\gedfn$-Walk and $\gccfn$ maximization.}
\label{fig:run_time_ged_stoch}
\end{subfigure}%
\hfill
\begin{subfigure}[t]{.5\columnwidth}
\includegraphics{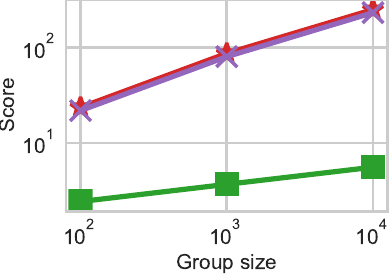}
\caption{$\gedfn$-Walk and group-closeness scores of groups computed by $\gedfn$
and $\gccfn$ maximization.
For each $k$ scores are divided by the scores for $k = 1$.}
\label{fig:scores}
\end{subfigure}
\caption{Running time (s) and scores of $\gedfn$-Walk with lazy-greedy ($\gedfn$) and stochastic greedy
($\gedfn$-S) strategies with large $k$ (log-scale).
Data points are aggregated using the geometric mean over the instances in Table~\ref{tab:running_time}.}
\end{figure}

Following the approach of~\cite{mirzasoleiman2015lazier},
for large values of $k$,
we can further improve the running
time performance of the greedy algorithm by using random sampling. This algorithm
is similar to the lazy algorithm. However, instead of considering
all vertices when maximizing the marginal gain (or when trying to
$(\epsilon/k)$-separate the vertex with maximal gain from the others),
\emph{stochastic greedy} only considers a subset of all vertices.
In particular, whenever the algorithm needs to add a new vertex
to the group, it samples $(\frac{|V|}k \log 1/\eta)$ vertices.
Maximization of the marginal gain and separation only consider
those vertices and ignore all others
(\ie only the sampled vertices are inserted into the
priority queues).
Note that if $(\epsilon/k)$-separation fails, the sampled vertices
are not discarded; the algorithm does exactly one round
of sampling per addition to the group.
This is necessary as the probability that we find a vertex
with high marginal gain is not independent from the
probability that the vertex can be separated against
other vertices of the sample.

\paragraph{Complexity of Stochastic Algorithm.}
In contrast to the lazy greedy algorithm, the stochastic greedy algorithm
evaluates $\gedfn$
at most $(\frac{|V|}k \log 1/\eta)$ times per $(\epsilon/k)$-separation, instead of $|V|$ times.
Hence, it has a worst-case complexity of
$\mathcal{O}(|V| (\log 1/\eta) \frac{\log (k |V|/\epsilon)}{\log 1/(\alpha \sigma_\mathrm{max})}
	(|V|+|E|))$.
In contrast to the lazy greedy algorithm, the relative approximation
quality of the stochastic greedy algorithm is $(1 - \frac 1e - \eta)$
instead of $(1 - \frac 1e)$.

\paragraph{Experiments on Large $k$.}
We provide additional experimental data including also the stochastic algorithm
with $\epsilon = 0.5$ and $\eta = 0.1$:
Figure~\ref{fig:run_time_ged_stoch} shows the running time of $\gedfn$-Walk
(both lazy and stochastic greedy strategies) and
$\gccfn$ maximization for large values of $k$.
For $k \ge \numprint{1000}$, the stochastic algorithm is
significantly faster than the lazy algorithm.
However, as explained in Section~\ref{sub:scalability_k},
$\gccfn$ maximization scales better than both $\gedfn$ strategies \wrt $k$, and for
$k \ge \numprint{1000}$ it is even faster than both $\gedfn$ and $\gedfn$-S.
On the other hand, Figure~\ref{fig:scores} shows the relative $\gedfn$-Walk and group-closeness
scores of the groups computed using $\gedfn$, $\gedfn$-S, and $\gccfn$.
The figure demonstrates that the stochastic greedy approach computes groups with the same
quality as $\gedfn$ in less time, which makes it a reasonable alternative to $\gedfn$ for
large values of $k$.

\section{GED-Walk on Large Real-World Graphs}
\label{apx:additional_exp}
Table~\ref{tab:large_inst_table} includes additional experimental data of $\gedfn$-Walk maximization
on large real-world networks with $k = 10$.
Results show that, for small group sizes, it is possible to maximize $\gedfn$-Walk on networks with hundreds
of millions of edges in a matter of minutes.

\begin{table*}[tb]
\centering
\footnotesize
\caption{Running time (s) of GED-Walk maximization on 36 cores on large real-world networks, $k = 10$.}
\label{tab:large_inst_table}
\centering
\begin{tabular}{lrrrr}
\toprule
Network & Category & $|V|$ & $|E|$ & Time (s) \\
\hline
petster-friendships-cat & Social & \numprint{148826} & \numprint{5447464} & \numprint{4.7}\\
dimacs9-W & Road & \numprint{6262104} & \numprint{7559642} & \numprint{45.3}\\
dimacs9-CTR & Road & \numprint{14081816} & \numprint{16933413} & \numprint{86.9}\\
flickr-growth & Social & \numprint{2173370} & \numprint{22729227} & \numprint{47.2}\\
soc-LiveJournal1 & Social & \numprint{4843953} & \numprint{42845684} & \numprint{35.0}\\
livejournal-links & Social & \numprint{5189808} & \numprint{48687945} & \numprint{47.1}\\
orkut-links & Social & \numprint{3072441} & \numprint{117184899} & \numprint{76.5}\\
dbpedia-link & Hyperlink & \numprint{18265512} & \numprint{126888089} & \numprint{348.7}\\
dimacs10-uk-2002 & Hyperlink & \numprint{18459128} & \numprint{261556721} & \numprint{47.5}\\
wikipedia\_link\_en & Hyperlink & \numprint{13591759} & \numprint{334590793} & \numprint{276.4}\\
\toprule
\end{tabular}
\end{table*}

\section{Impact of Parameter $\alpha$}
\label{apx:impact_of_alpha}
\begin{figure}[t]
\centering
\begin{subfigure}[t]{\columnwidth}
\centering
\includegraphics{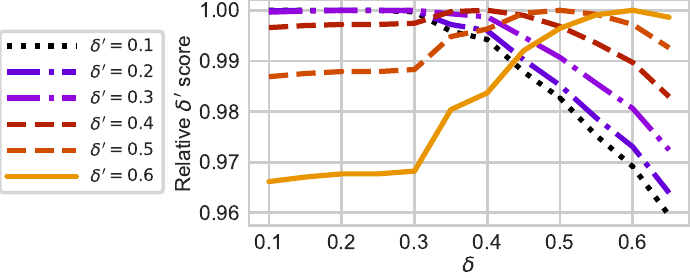}
\caption{Relative $\delta'$ score
	(\ie $\gedfn^{\delta'}(S^\delta) / \gedfn^{\delta}(S^\delta)$)
for $\delta, \delta' \in[0.1, \dots, 0.6]$, $\epsilon = 0.1$, and $k = 10$. }
\label{fig:relative_score_delta}
\end{subfigure}
\smallskip\par
\begin{subfigure}[t]{\columnwidth}
\centering
\includegraphics{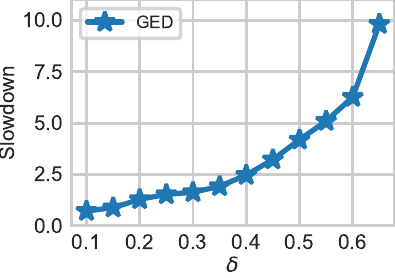}
\caption{Slowdown of our algorithm for maximizing GED-Walk using the spectral bound
with $\alpha = \delta / \sigma_{\mathrm{max}}$
over the lazy algorithm using the combinatorial bound.}
\label{fig:run_time_sdelta}
\end{subfigure}
\caption{Quality and running time performance of  using the spectral bound.
Data points are aggregated using the geometric mean over the instances in Table~\ref{tab:rt_instances_table}.}
\label{fig:spectral_delta}
\end{figure}

Finally, we analyze how different settings of the parameter $\alpha$ impact
our the resulting $\gedfn$ groups.
Here, we use the spectral bound (see Equation~\ref{eq:spectral-bound}).
In Proposition~\ref{prop:katz-relation} we show that GED-Walk converges
iff $\alpha < 1/\sigma_{\mathrm{max}}$.
Let $\delta \in (0,1]$; obviously, GED-Walk also converges if $\alpha < \delta/\sigma_{\mathrm{max}}$.
In this experiment we compute groups $S^\delta$ for a certain value $\alpha = \delta / \sigma_{\mathrm{max}}$;
we measure the score of the resulting group for $\alpha' = \delta' / \sigma_{\mathrm{max}}$.
Denote this score by $\gedfn^{\delta'}(S^\delta)$.
Figure~\ref{fig:relative_score_delta} shows the ratio
$\gedfn^{\delta'}(S^\delta) / \gedfn^{\delta}(S^\delta)$,
with $\epsilon = 0.1$, $\delta, \delta' \in \{0.1, \dots, 0.6\}$, and $k = 10$.
Figure~\ref{fig:run_time_sdelta} shows the slowdown \ie running time of computing
$S^\delta$ divided by the running time of the lazy algorithm using the combinatorial bound.
Computing $\gedfn^{\delta'}(S^\delta)$ with $\delta \in[0.1, \dots, 0.3]$
yields similar scores independently of
$\delta$, which means that
the GED-Walk score of a group does not change significantly in such a $\delta$ interval.
Increasing $\delta$ above $0.3$ leads to a noticeable reduction of
the relative scores computed
using $\delta' \le 0.3$, and to a steeper growth of the ones computed using $\delta' \ge 0.5$.

\begin{table*}
\footnotesize
\caption{Running time (s) of $\gbcfn$, $\gccfn$, and $\gedfn$ maximization on 36 cores for different
group sizes ($5$, $10$, $20$, $50$, and $100$).}
\label{tab:running_time}
\centering
\begin{tabular}{lrrrrrrrrrr}
\toprule
\multirow{3}{*}{Network} & \multicolumn{2}{c}{$k = 5$} & \multicolumn{2}{c}{$k = 10$} & \multicolumn{2}{c}{$k = 20$} & \multicolumn{2}{c}{$k = 50$} & \multicolumn{2}{c}{$k = 100$}\\
& $\gbcfn$& $\gccfn$& $\gbcfn$& $\gccfn$& $\gbcfn$& $\gccfn$& $\gbcfn$& $\gccfn$& $\gbcfn$& $\gccfn$\\
& $\gedfn$& $\gedfn$-S& $\gedfn$& $\gedfn$-S& $\gedfn$& $\gedfn$-S& $\gedfn$& $\gedfn$-S& $\gedfn$& $\gedfn$-S\\

\hline
\multirow{2}{*}{dimacs9-COL} & \numprint{21.3} & \numprint{454.0} & \numprint{42.0} & \numprint{480.6} & \numprint{84.2} & \numprint{529.5} & \numprint{207.5} & \numprint{578.7} & \numprint{412.6} & \numprint{616.8}\\
 & \numprint{1.0} & \numprint{1.0} & \numprint{1.4} & \numprint{1.2} & \numprint{2.5} & \numprint{1.9} & \numprint{4.4} & \numprint{4.2} & \numprint{6.8} & \numprint{38.0}\\

\hline
\multirow{2}{*}{munmun\_twitter\_social} & \numprint{48.8} & \numprint{20.1} & \numprint{95.7} & \numprint{21.4} & \numprint{189.9} & \numprint{24.9} & \numprint{475.0} & \numprint{36.4} & \numprint{952.1} & \numprint{42.6}\\
 & \numprint{0.7} & \numprint{0.7} & \numprint{1.0} & \numprint{0.9} & \numprint{1.5} & \numprint{1.5} & \numprint{3.8} & \numprint{3.4} & \numprint{9.1} & \numprint{6.4}\\

\hline
\multirow{2}{*}{com-dblp} & \numprint{36.7} & \numprint{29.4} & \numprint{71.7} & \numprint{30.3} & \numprint{141.1} & \numprint{33.0} & \numprint{351.7} & \numprint{41.0} & \numprint{709.2} & \numprint{48.3}\\
 & \numprint{1.2} & \numprint{1.1} & \numprint{1.5} & \numprint{1.4} & \numprint{2.5} & \numprint{2.4} & \numprint{8.3} & \numprint{5.5} & \numprint{16.4} & \numprint{11.4}\\

\hline
\multirow{2}{*}{wikipedia\_link\_mr} & \numprint{17.7} & \numprint{0.9} & \numprint{35.3} & \numprint{1.0} & \numprint{70.2} & \numprint{1.1} & \numprint{175.1} & \numprint{1.5} & \numprint{354.3} & \numprint{2.0}\\
 & \numprint{0.3} & \numprint{0.3} & \numprint{0.5} & \numprint{0.6} & \numprint{0.9} & \numprint{0.8} & \numprint{2.0} & \numprint{1.8} & \numprint{4.6} & \numprint{4.0}\\

\hline
\multirow{2}{*}{roadNet-PA} & \numprint{66.3} & \numprint{3903.0} & \numprint{131.9} & \numprint{4150.4} & \numprint{261.2} & \numprint{4353.8} & \numprint{645.5} & \numprint{4676.0} & \numprint{1292.3} & \numprint{4911.8}\\
 & \numprint{3.2} & \numprint{3.0} & \numprint{4.1} & \numprint{4.1} & \numprint{7.3} & \numprint{5.2} & \numprint{11.5} & \numprint{13.4} & \numprint{28.3} & \numprint{30.0}\\

\hline
\multirow{2}{*}{citeseer} & \numprint{74.3} & \numprint{52.5} & \numprint{146.4} & \numprint{56.9} & \numprint{290.0} & \numprint{62.5} & \numprint{717.9} & \numprint{76.5} & \numprint{1440.2} & \numprint{87.7}\\
 & \numprint{1.1} & \numprint{1.1} & \numprint{1.4} & \numprint{1.5} & \numprint{2.6} & \numprint{2.3} & \numprint{5.5} & \numprint{5.8} & \numprint{11.8} & \numprint{10.1}\\

\hline
\multirow{2}{*}{roadNet-TX} & \numprint{83.9} & \numprint{4903.0} & \numprint{165.3} & \numprint{5230.8} & \numprint{325.6} & \numprint{5675.7} & \numprint{814.9} & \numprint{6149.2} & \numprint{1634.9} & \numprint{6485.9}\\
 & \numprint{2.4} & \numprint{2.4} & \numprint{3.4} & \numprint{3.2} & \numprint{4.9} & \numprint{4.6} & \numprint{9.4} & \numprint{14.8} & \numprint{28.4} & \numprint{26.4}\\

\hline
\multirow{2}{*}{web-Stanford} & \numprint{21.7} & \numprint{57.8} & \numprint{43.1} & \numprint{59.8} & \numprint{88.0} & \numprint{63.2} & \numprint{223.8} & \numprint{63.0} & \numprint{450.6} & \numprint{66.1}\\
 & \numprint{0.7} & \numprint{0.7} & \numprint{1.1} & \numprint{1.1} & \numprint{3.2} & \numprint{2.1} & \numprint{6.6} & \numprint{4.2} & \numprint{12.8} & \numprint{8.1}\\

\hline
\multirow{2}{*}{petster-dog-household} & \numprint{83.2} & \numprint{17.3} & \numprint{164.9} & \numprint{17.4} & \numprint{328.1} & \numprint{17.8} & \numprint{823.5} & \numprint{19.8} & \numprint{1652.9} & \numprint{24.1}\\
 & \numprint{0.4} & \numprint{0.4} & \numprint{0.5} & \numprint{0.6} & \numprint{0.8} & \numprint{0.9} & \numprint{1.9} & \numprint{1.8} & \numprint{4.0} & \numprint{3.5}\\

\hline
\multirow{2}{*}{wikipedia\_link\_bn} & \numprint{41.0} & \numprint{6.0} & \numprint{81.5} & \numprint{6.2} & \numprint{162.7} & \numprint{7.2} & \numprint{406.3} & \numprint{8.8} & \numprint{815.9} & \numprint{10.7}\\
 & \numprint{0.7} & \numprint{0.7} & \numprint{1.1} & \numprint{1.1} & \numprint{1.7} & \numprint{1.7} & \numprint{3.9} & \numprint{3.6} & \numprint{8.4} & \numprint{6.5}\\

\hline
\multirow{2}{*}{petster-catdog-household} & \numprint{95.6} & \numprint{61.5} & \numprint{189.0} & \numprint{61.8} & \numprint{376.2} & \numprint{62.3} & \numprint{935.6} & \numprint{64.4} & \numprint{1874.0} & \numprint{70.2}\\
 & \numprint{0.5} & \numprint{0.6} & \numprint{0.7} & \numprint{0.8} & \numprint{1.4} & \numprint{1.5} & \numprint{3.2} & \numprint{3.2} & \numprint{6.4} & \numprint{6.0}\\

\hline
\multirow{2}{*}{wikipedia\_link\_uz} & \numprint{46.1} & \numprint{20.7} & \numprint{91.4} & \numprint{21.3} & \numprint{182.5} & \numprint{22.0} & \numprint{461.0} & \numprint{25.2} & \numprint{932.3} & \numprint{31.1}\\
 & \numprint{0.7} & \numprint{0.8} & \numprint{1.1} & \numprint{1.1} & \numprint{1.7} & \numprint{1.9} & \numprint{4.2} & \numprint{4.3} & \numprint{9.1} & \numprint{9.5}\\
\toprule
\end{tabular}
\end{table*}

\begin{table*}
    \centering
    \footnotesize
    \caption{Graph classification accuracy(in \%). \textsc{PR} denotes PageRank with all vertices as the teleport set.}
	\begin{tabular}{lrrrrr}
		\toprule
		Dataset &  ENZYMES &  IMDB-BINARY &  Mutagenicity &  PROTEINS &  REDDIT-BINARY \\
												 &          &              &               &           &                \\
		\midrule
		\textsc{Eig}-H+\textsc{Ged+PPR}$^*$-H    &    24.12 &\textbf{71.62}& \textbf{69.18}&     73.10 &          74.17 \\
		\textsc{Eig}-T+\textsc{Ged+PPR}$^*$-T    &    27.45 &        69.25 &         62.78 &     73.54 &          76.70 \\
		\textsc{Eig}-H+\textsc{Ged+PPR}$^{**}$-H &    24.78 &        70.54 &         68.81 &     72.97 &  \textbf{81.43} \\
		\textsc{Eig}-T+\textsc{Ged+PPR}$^{**}$-T &\textbf{27.88} &   68.53 &         62.43 &     73.72 &          80.48 \\
		\textsc{Eig}-H+\textsc{Ged}              &    23.14 &        69.74 &         69.08 &     73.13 &          75.16 \\
		\textsc{Eig}-T+\textsc{Ged}              &    26.46 &        63.56 &         64.14 &\textbf{74.06} &      80.18 \\
		\textsc{Eig}-H+\textsc{PPR}$^*$-H        &    24.20 &\textbf{71.62}& \textbf{69.18}&     73.08 &          74.17 \\
		\textsc{Eig}-T+\textsc{PPR}$^*$-T        &    27.49 &        69.25 &         62.78 &     73.56 &          76.71 \\
		\textsc{Eig}-H+\textsc{PPR}$^{**}$-H     &    24.78 &        70.54 &         68.80 &     72.97 &  \textbf{81.43} \\
		\textsc{Eig}-T+\textsc{PPR}$^{**}$-T     &\textbf{27.88} &   68.53 &         62.43 &     73.72 &          80.48 \\
		\textsc{Eig}-H+\textsc{PR}$^*$-H         &    23.54 &        70.00 &         68.85 &     72.33 &          75.11 \\
		\textsc{Eig}-T+\textsc{PR}$^*$-T         &    23.18 &        57.06 &         57.38 &     73.25 &          76.00 \\
		\textsc{Eig}-H+\textsc{PR}$^{**}$-H      &    23.54 &        70.42 &         69.17 &     72.49 &          76.33 \\
		\textsc{Eig}-T+\textsc{PR}$^{**}$-T      &    23.18 &        58.42 &         57.79 &     73.31 &          76.59 \\
		\textsc{Eig}-H                           &    23.47 &        70.28 &         68.88 &     72.42 &          72.02 \\
		\textsc{Eig}-T                           &    23.02 &        56.59 &         56.90 &     73.35 &          75.31 \\
		\textsc{Ged+PPR}$^*$-H                   &    20.85 &        65.18 &         65.46 &     72.18 &          71.59 \\
		\textsc{Ged+PPR}$^{**}$-H                &    20.39 &        66.27 &         65.86 &     72.44 &          75.95 \\
		\textsc{Ged}                             &    19.18 &        60.64 &         64.51 &     71.95 &          70.59 \\
		\textsc{PPR}$^*$-H                       &    21.01 &        60.93 &         60.44 &     71.79 &          72.74 \\
		\textsc{PPR}$^{**}$-H                    &    19.46 &        66.39 &         62.27 &     71.86 &          73.79 \\
		\textsc{PR}$^*$-H                        &    20.11 &        57.04 &         61.31 &     71.78 &          73.19 \\
		\textsc{PR}$^*$-T                      &    16.98 &        51.15 &         55.85 &     71.84 &          69.66 \\
		\textsc{PR}$^{**}$-H                     &    20.11 &        60.84 &         61.75 &     71.77 &          73.43 \\
		\textsc{PR}$^{**}$-T                   &    16.98 &        56.64 &         57.19 &     72.12 &          73.31 \\
		\bottomrule
		\multicolumn{6}{l}{$^*$: PageRank teleport probability 0.85; $^{**}$: PageRank teleport probability 0.15}
		\end{tabular}
		\label{tab:all_graph_classification}
\end{table*}

\end{document}